%% file: pr_p11d13_070212.tex
\documentclass[prc,twocolumn,superscriptaddress,showpacs,amssymb,amsmath,amsfonts,aps]{revtex4}
\usepackage{graphicx}
\usepackage{dcolumn}
\usepackage{epsfig}
\usepackage{latexsym}
\usepackage{amsmath} 
\usepackage[dvips]{color}

\begin{document}

\date{\today}
\title{\large A study of the $P_{11}(1440)$ and $D_{13}(1520)$ resonances from 
CLAS data on $ep \rightarrow e'\pi^{+} \pi^{-} p'$  \\ }
\newcommand*{\JLAB}{Thomas Jefferson National Accelerator Facility, Newport News, Virginia 23606}
\newcommand*{\JLABindex}{33}
\affiliation{\JLAB}
\newcommand*{\MSU}{Skobeltsyn Nuclear Physics Institute and Physics Department at Moscow State University, 119899 Moscow, Russia}
\newcommand*{\MSUindex}{31}
\affiliation{\MSU}
\newcommand*{\SCAROLINA}{University of South Carolina, Columbia, South Carolina 29208}
\newcommand*{\SCAROLINAindex}{32}
\affiliation{\SCAROLINA}
\newcommand*{\ANL}{Argonne National Laboratory, Argonne, Illinois 60439}
\newcommand*{\ANLindex}{1}
\affiliation{\ANL}
\newcommand*{\ASU}{Arizona State University, Tempe, Arizona 85287-1504}
\newcommand*{\ASUindex}{2}
\affiliation{\ASU}
\newcommand*{\CSUDH}{California State University, Dominguez Hills, Carson, CA 90747}
\newcommand*{\CSUDHindex}{3}
\affiliation{\CSUDH}
\newcommand*{\CANISIUS}{Canisius College, Buffalo, NY}
\newcommand*{\CANISIUSindex}{4}
\affiliation{\CANISIUS}
\newcommand*{\CMU}{Carnegie Mellon University, Pittsburgh, Pennsylvania 15213}
\newcommand*{\CMUindex}{5}
\affiliation{\CMU}
\newcommand*{\CUA}{Catholic University of America, Washington, D.C. 20064}
\newcommand*{\CUAindex}{6}
\affiliation{\CUA}
\newcommand*{\SACLAY}{CEA, Centre de Saclay, Irfu/Service de Physique Nucl\'eaire, 91191 Gif-sur-Yvette, France}
\newcommand*{\SACLAYindex}{7}
\affiliation{\SACLAY}
\newcommand*{\CNU}{Christopher Newport University, Newport News, Virginia 23606}
\newcommand*{\CNUindex}{8}
\affiliation{\CNU}
\newcommand*{\UCONN}{University of Connecticut, Storrs, Connecticut 06269}
\newcommand*{\UCONNindex}{9}
\affiliation{\UCONN}
\newcommand*{\FU}{Fairfield University, Fairfield CT 06824}
\newcommand*{\FUindex}{10}
\affiliation{\FU}
\newcommand*{\FIU}{Florida International University, Miami, Florida 33199}
\newcommand*{\FIUindex}{11}
\affiliation{\FIU}
\newcommand*{\FSU}{Florida State University, Tallahassee, Florida 32306}
\newcommand*{\FSUindex}{12}
\affiliation{\FSU}
\newcommand*{\Genova}{Universit$\grave{a}$ di Genova, 16146 Genova, Italy}
\newcommand*{\Genovaindex}{13}
\affiliation{\Genova}
\newcommand*{\GWUI}{The George Washington University, Washington, DC 20052}
\newcommand*{\GWUIindex}{14}
\affiliation{\GWUI}
\newcommand*{\ISU}{Idaho State University, Pocatello, Idaho 83209}
\newcommand*{\ISUindex}{15}
\affiliation{\ISU}
\newcommand*{\INFNFE}{INFN, Sezione di Ferrara, 44100 Ferrara, Italy}
\newcommand*{\INFNFEindex}{16}
\affiliation{\INFNFE}
\newcommand*{\INFNFR}{INFN, Laboratori Nazionali di Frascati, 00044 Frascati, Italy}
\newcommand*{\INFNFRindex}{17}
\affiliation{\INFNFR}
\newcommand*{\INFNGE}{INFN, Sezione di Genova, 16146 Genova, Italy}
\newcommand*{\INFNGEindex}{18}
\affiliation{\INFNGE}
\newcommand*{\INFNRO}{INFN, Sezione di Roma Tor Vergata, 00133 Rome, Italy}
\newcommand*{\INFNROindex}{19}
\affiliation{\INFNRO}
\newcommand*{\ORSAY}{Institut de Physique Nucl\'eaire ORSAY, Orsay, France}
\newcommand*{\ORSAYindex}{20}
\affiliation{\ORSAY}
\newcommand*{\ITEP}{Institute of Theoretical and Experimental Physics, Moscow, 117259, Russia}
\newcommand*{\ITEPindex}{21}
\affiliation{\ITEP}
\newcommand*{\JMU}{James Madison University, Harrisonburg, Virginia 22807}
\newcommand*{\JMUindex}{22}
\affiliation{\JMU}
\newcommand*{\KNU}{Kyungpook National University, Daegu 702-701, Republic of Korea}
\newcommand*{\KNUindex}{23}
\affiliation{\KNU}
\newcommand*{\LPSC}{LPSC, Universite Joseph Fourier, CNRS/IN2P3, INPG, Grenoble, France
}
\newcommand*{\LPSCindex}{24}
\affiliation{\LPSC}
\newcommand*{\UNH}{University of New Hampshire, Durham, New Hampshire 03824-3568}
\newcommand*{\UNHindex}{25}
\affiliation{\UNH}
\newcommand*{\NSU}{Norfolk State University, Norfolk, Virginia 23504}
\newcommand*{\NSUindex}{26}
\affiliation{\NSU}
\newcommand*{\OHIOU}{Ohio University, Athens, Ohio  45701}
\newcommand*{\OHIOUindex}{27}
\affiliation{\OHIOU}
\newcommand*{\ODU}{Old Dominion University, Norfolk, Virginia 23529}
\newcommand*{\ODUindex}{28}
\affiliation{\ODU}
\newcommand*{\RPI}{Rensselaer Polytechnic Institute, Troy, New York 12180-3590}
\newcommand*{\RPIindex}{29}
\affiliation{\RPI}
\newcommand*{\UNIONC}{Union College, Schenectady, NY 12308}
\newcommand*{\UNIONCindex}{30}
\affiliation{\UNIONC}
\newcommand*{\URICH}{University of Richmond, Richmond, Virginia 23173}
\newcommand*{\URICHindex}{31}
\affiliation{\URICH}
\newcommand*{\ROMAII}{Universita' di Roma Tor Vergata, 00133 Rome Italy}
\newcommand*{\ROMAIIindex}{32}
\affiliation{\ROMAII}

\newcommand*{\UTFSM}{Universidad T\'{e}cnica Federico Santa Mar\'{i}a, Casilla 110-V Valpara\'{i}so, Chile}
\newcommand*{\UTFSMindex}{33}
\affiliation{\UTFSM}
\newcommand*{\GLASGOW}{University of Glasgow, Glasgow G12 8QQ, United Kingdom}
\newcommand*{\GLASGOWindex}{34}
\affiliation{\GLASGOW}
\newcommand*{\VIRGINIA}{University of Virginia, Charlottesville, Virginia 22901}
\newcommand*{\VIRGINIAindex}{35}
\affiliation{\VIRGINIA}
\newcommand*{\WM}{College of William and Mary, Williamsburg, Virginia 23187-8795}
\newcommand*{\WMindex}{36}
\affiliation{\WM}
\newcommand*{\YEREVAN}{Yerevan Physics Institute, 375036 Yerevan, Armenia}
\newcommand*{\YEREVANindex}{37}
\affiliation{\YEREVAN}

\newcommand*{\NOWSACLAY}{CEA, Centre de Saclay, Irfu/Service de Physique Nucl\'eaire, 91191 Gif-sur-Yvette, France}
\newcommand*{\NOWMSU}{Skobeltsyn Nuclear Physics Institute, 119899 Moscow, Russia}
\newcommand*{\NOWORSAY}{Institut de Physique Nucl\'eaire ORSAY, Orsay, France}
\newcommand*{\NOWINFNGE}{INFN, Sezione di Genova, 16146 Genova, Italy}
\newcommand*{\NOWROMAII}{Universita' di Roma Tor Vergata, 00133 Rome Italy}

\author {V.I.~Mokeev} 
\altaffiliation[Current address:]{\JLAB}
\affiliation{\JLAB}
\affiliation{\MSU}
\author {V.D.~Burkert} 
\affiliation{\JLAB}
\author {L.~Elouadrhiri} 
\affiliation{\JLAB}
\author {G.V.~Fedotov} 
\affiliation{\SCAROLINA}
\affiliation{\MSU}
\author {E.N.~Golovatch} 
\affiliation{\MSU}
\author {R.W.~Gothe} 
\affiliation{\SCAROLINA}
\author {B.S.~Ishkhanov} 
\affiliation{\MSU}
\author {E.L.~Isupov} 
\affiliation{\MSU}

\author {K.P. ~Adhikari} 
\affiliation{\ODU}
\author {M.~Aghasyan} 
\affiliation{\INFNFR}
% ...

\collaboration{The CLAS Collaboration} \noaffiliation



\begin{abstract}
{ The transition helicity amplitudes from the proton ground state to the  $P_{11}(1440)$ and  $D_{13}(1520)$  excited states
($\gamma_{v}pN^*$ electrocouplings) were determined from the analysis of nine independent
one-fold differential $\pi^{+} \pi^{-} p$ electroproduction cross sections
 off a proton target, taken with CLAS at photon virtualities
0.25\enskip {\rm GeV$^{2}$} $<$ $Q^{2}$ $<$ 0.60 \enskip {\rm GeV$^{2}$}. The phenomenological reaction model was employed for separation of the resonant and non-resonant contributions to the final state. 
The $P_{11}(1440)$ and $D_{13}(1520)$ electrocouplings were obtained from the resonant amplitudes parametrized within the framework of a unitarized Breit-Wigner ansatz. 
They  
are consistent with results obtained in the previous CLAS analyses of the 
$\pi^+n$ and  $\pi^0p$ channels. The successful
description  of a large body of data  
in dominant meson-electroproduction channels off protons with the same $\gamma_{v}pN^*$ 
electrocouplings offers clear evidence 
for the reliable extraction of these fundamental quantities from  
meson-electroproduction data. This analysis also led to the determination 
of the 
long-awaited 
hadronic branching ratios for the $D_{13}(1520)$ decay into $\Delta\pi$ (24\%-32\%) and 
$N\rho$ (8\%-17\%). } 
\end{abstract}


\maketitle

\section{Introduction}

An extensive research program on nucleon resonance ($N^*$) excitation 
is in progress using the CLAS detector in
Hall-B at Jefferson Lab \cite{Bu11a,Bu11b,Bu11,Bu05,Bu04}. The studies of transition helicity amplitudes 
from the proton ground state to its excited states (or $\gamma_{v}pN^*$ photo-/electrocouplings) represent a key
direction in the $N^*$ program with CLAS. Meson-electroproduction data off nucleons in the $N^*$ region 
obtained
with CLAS open up an opportunity to determine
the $Q^2$-evolution of 
$\gamma_{v}NN^*$ electrocouplings in a combined analysis of various meson-electroproduction channels.
The $Q^2$-evolution of $\gamma_{v}NN^*$ electrocouplings 
will allow us to pin down active
degrees of freedom in the $N^*$ structure at various distance scales 
and to access non-perturbative 
strong-interaction mechanisms that govern 
the excited nucleon state formation as bound systems of quarks and gluons.

Theoretical and experimental studies
of the electroexcitation of nucleon resonances
have a long history. 
Along with the hadron masses and their partial decay widths, the information on the
$\gamma_{v}pN^*$ electrocouplings played
an important role in the development  of the quark models in their contemporary advanced relativistic version
in light-front dynamics \cite{Brodsky,Aznquark1,Capstick,Simula1,Bruno,AznRoper,AzBu12}. 
The picture of the nucleon 
and its excited states,
which seemed quite simply modeled with three relativistic 
constituent quarks,
turned out to be more complex. Recently obtained electrocouplings for the
$P_{11}(1440)$, $D_{13}(1520)$, and 
$S_{11}(1535)$ states \cite{Az09} showed that those quark models, which were 
successful in describing
the electrocouplings of these states for $Q^2$ $>$ 2.0  GeV$^2$, failed to reproduce the results 
at smaller photon virtualities of $Q^2$ $<$ 1.0 GeV$^2$. Quark models with flavor-conserving quark interactions are unable to describe 
the small mass of the $P_{11}(1440)$ state \cite{Ca00} and the ordering of $P_{11}(1440)$ and
$S_{11}(1535)$ resonances \cite{Bu11b}. Moreover, models that treat the $P_{11}(1440)$ structure as just three constituent quarks 
are unable 
to describe the large total $P_{11}(1440)$ decay width of $\approx$ 300 MeV. 
These
difficulties prompted a search for additional contributions to the $N^*$ structure.

A general unitarity requirement imposes meson-baryon dressing contributions to
both resonance electromagnetic excitation and hadronic decay amplitudes.
Studies of meson-baryon dressing contributions to $\gamma_{v}pN^*$ 
electrocouplings 
and resonance hadronic decay
amplitudes, carried out at the Excited Baryon Analysis Center (EBAC) at 
Jefferson Lab \cite{Lee10,Lee09,Lee091,Lee08,Lee101,Lee101a,Lee102,Lee11c}, have extended our insight into the spectrum and structure of excited nucleon states 
considerably. The contributions from 
meson-baryon dressing to the $P_{11}(1440)$ and $D_{13}(1520)$ electrocouplings
were determined 
from a global analysis of the world data on $\pi N$ scattering and $\pi^+n$, $\pi^0p$
electroproduction off protons within the framework of the EBAC dynamical 
coupled-channel 
approach (EBAC-DCC) \cite{Lee10}. 

This analysis showed that the contributions from meson-baryon dressing 
to the $\gamma_{v}pN^*$ electrocouplings are maximal at
small $Q^2$ and decrease with increasing photon virtualities \cite{Lee08}.
At $Q^2$ $<$ 1.0 GeV$^2$ these contributions
may even be dominant. The meson-baryon cloud 
has a profound impact on the resonance spectrum. 
For example, in the $P_{11}$ partial wave 
a single bare resonance pole 
located at $W$=1.76 GeV, being affected by meson-baryon dressing, splits into 
three poles located on different Riemann sheets \cite{Lee101,Lee102}. Two of them  
with $Re(W)$ $\approx$ 1.36 GeV correspond to the physical Roper resonance. The double-pole structure of the Roper resonance was also observed in previous studies \cite{Ar85,Ar06}. Meson-baryon dressing should be taken
into consideration in the interpretation of resonance electrocouplings 
as well as for the
excited nucleon spectrum.

Our studies of resonance electrocouplings 
at small photon
virtualities presented in this paper offer valuable information to further 
explore
the role of meson-baryon and quark components in the $N^*$ structure. The separation between the meson-baryon cloud and quark core contributions within a well-defined 
theoretical
framework \cite{Ca07} can help pin down the domain of photon virtualities
where  quark components are the main contributor to the $N^*$
structure. This kinematic domain is of particular interest for the studies of hadrons from the first principles of QCD, including Lattice QCD (LQCD) \cite{Br09,Li09,Li09a,Li11,Li11a,Ed11,Ed12,CL12} and Dyson-Schwinger equation studies of 
QCD (DSEQCD) \cite{CL12,CRo09,CRo011,CRo101,CRo103,CRo104,CR2012}.


The CLAS detector at 
Jefferson Lab
is a unique large-acceptance instrument
designed for the comprehensive exploration
of exclusive meson electroproduction. It offers excellent opportunities 
of studying the electroexcitation of nucleon
resonances in detail and with precision. The CLAS detector has provided the dominant portion of all data on meson electroproduction in the resonance excitation region.

A variety of measurements
of single pion electroproduction off protons,
including polarization measurements, have been performed
at CLAS in a range of $Q^2$ from 0.16 to 6 GeV$^2$ \cite{Joo}.
 The electroexcitation amplitudes
for the low-lying resonances
$P_{33}(1232)$, $P_{11}(1440)$, $D_{13}(1520)$,
and $S_{11}(1535)$ were determined over a wide range of $Q^2$
in a comprehensive analysis
of JLab-CLAS  data 
on differential cross sections, longitudinally polarized
beam asymmetries, and longitudinal target and beam-target asymmetries \cite{Az09}.

The combination of the large-acceptance CLAS detector and 
the continuous electron beam from CEBAF 
made it possible to measure $\pi^+\pi^-p$ electroproduction cross sections with
nearly full kinematic coverage for this three-body final hadron state \cite{Ri03,Fe09}. These are the most extensive data sets on unpolarized $\pi^+\pi^-p$ electroproduction cross sections obtained so far. These data allowed for the first time the projection  
of nine one-dimensional differential cross sections, each sensitive to  
a different combination of resonance and background strength. The data of \cite{Fe09}  were collected in the mass range 1.31 GeV $<$ $W$ $<$ 1.56 GeV and with
photon virtualities 0.25 GeV$^2$ $<$ $Q^2$ $<$ 0.6 GeV$^2$. A good description 
of these data was achieved
within the framework of a phenomenological Jefferson Laboratory - Moscow State University (JM) reaction model \cite{Mo09}, that
allowed 
us to establish the mechanisms contributing to this exclusive reaction. The presence and strength of the contributing
$\pi^+\pi^-p$ electroproduction mechanisms were established
by studying their kinematical dependencies and correlations in different 
one-fold differential cross sections.

In this work we present results on the electroexcitation of the $P_{11}(1440)$ and  $D_{13}(1520)$ states,
obtained from the analysis of data on $\pi^+\pi^-p$ electroproduction off protons \cite{Fe09}. The analysis was
carried out employing the JM reaction model \cite{Mo09}, which was further developed to provide a framework for 
the determination
of $\gamma_{v}pN^*$ electrocouplings from a combined fit of unpolarized differential cross sections. In the previous studies \cite{Mo09}  we did not attempt to isolate the
contributions from resonances. In the analysis reported in
this paper, we employ the JM model with the goal of isolating the resonant contributions 
for the individual differential cross sections. For the description of 
resonant amplitudes, 
we updated the Breit-Wigner (BW) parametrization, 
making
it consistent with the restrictions required by the general unitarity condition. The reliable evaluation of the resonant contributions enabled for the first time 
   the determination of the $\gamma_{v}pN^*$ electrocouplings for the $P_{11}(1440)$ and $D_{13}(1520)$ states 
from charged double pion
electroproduction off protons. 
This complements the results from $N\pi$ electroproduction in an 
independent channel. 

Analyses of different exclusive channels are essential for a 
reliable extraction of resonance parameters. Currently the separation of resonant and non-resonant 
parts of the electro production amplitudes can be done only within phenomenological 
reaction models. Therefore, the resonance parameters extracted from the meson electro production data fit may be affected by the model assumptions, and their credibility should be further examined. Non-resonant mechanisms in various meson-electroproduction channels are
completely different, while the $\gamma_{v}NN^*$ electrocouplings are the same.  Independent analyses of different exclusive channels make it possible to test whether they give consistent results for the
resonance electrocouplings. 
Most nucleon resonances decay into both $N\pi$ and $N\pi\pi$ final states. 
Studies of resonance electroexcitations 
in these channels with completely different non-resonant contributions offer independent information on $N^*$ electrocouplings.  
Therefore, a successful description of the data on $\pi^+n$, $\pi^0p$, and $\pi^+\pi^-p$ electroproduction off protons with 
consistent $N^*$
electrocoupling values provides clear evidence for the reliable extraction of these quantities 
from meson-electroproduction data.

Studies of the $P_{11}(1440)$ resonance 
in $\pi^+\pi^-p$ electroproduction off protons 
offer the additional opportunities to improve the knowledge on electrocouplings of this $\approx$ 300 MeV broad state. Contrary to $N\pi$
electroproduction channels, the $P_{33}(1232)$ resonance does not directly 
contribute to the resonant parts 
of the $\pi^+\pi^-p$ electroproduction amplitude. Hence, 
the influence of the $P_{33}(1232)$ resonance on the extracted electrocouplings of 
the $P_{11}(1440)$ is much weaker. 

The requirement of $Q^2$-independent $N^*$ hadronic decay amplitudes
 in  $\pi^+\pi^-p$ electroproduction  provides constraints on 
 the $N\pi\pi$ partial decay widths. This makes possible   
access to the $\pi\Delta$ and $\rho p$  partial hadronic decay widths of the $P_{11}(1440)$ and $D_{13}(1520)$ in the measurements of $\pi^+\pi^-p$ electroproduction off protons.

\begin{figure}[b]
\begin{center}
\includegraphics[width=8cm]{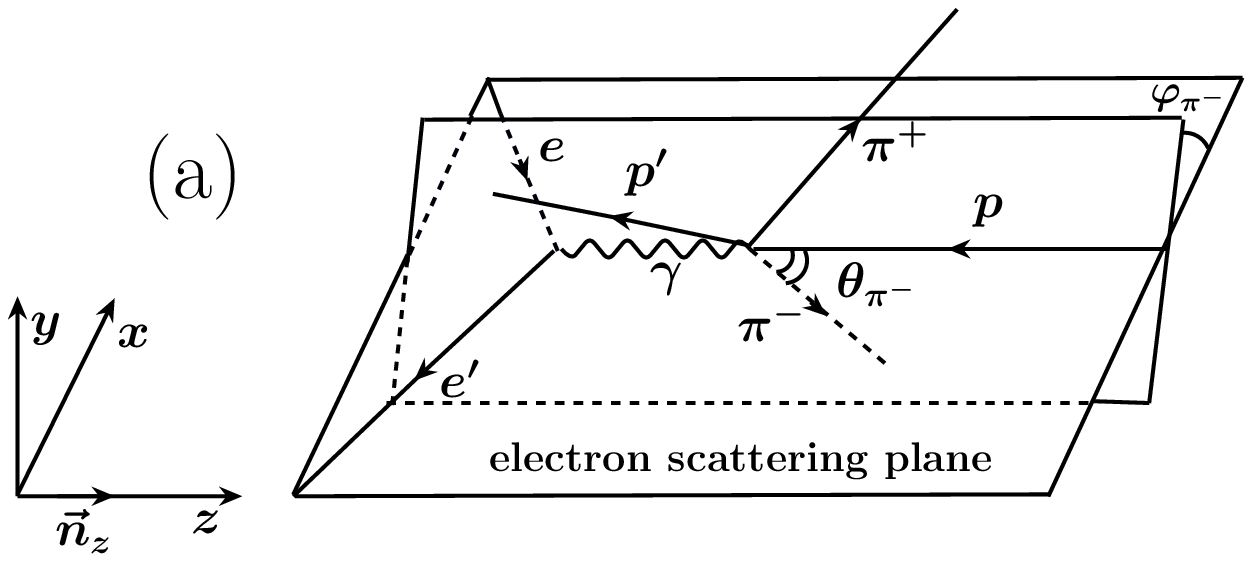}
\includegraphics[width=8cm]{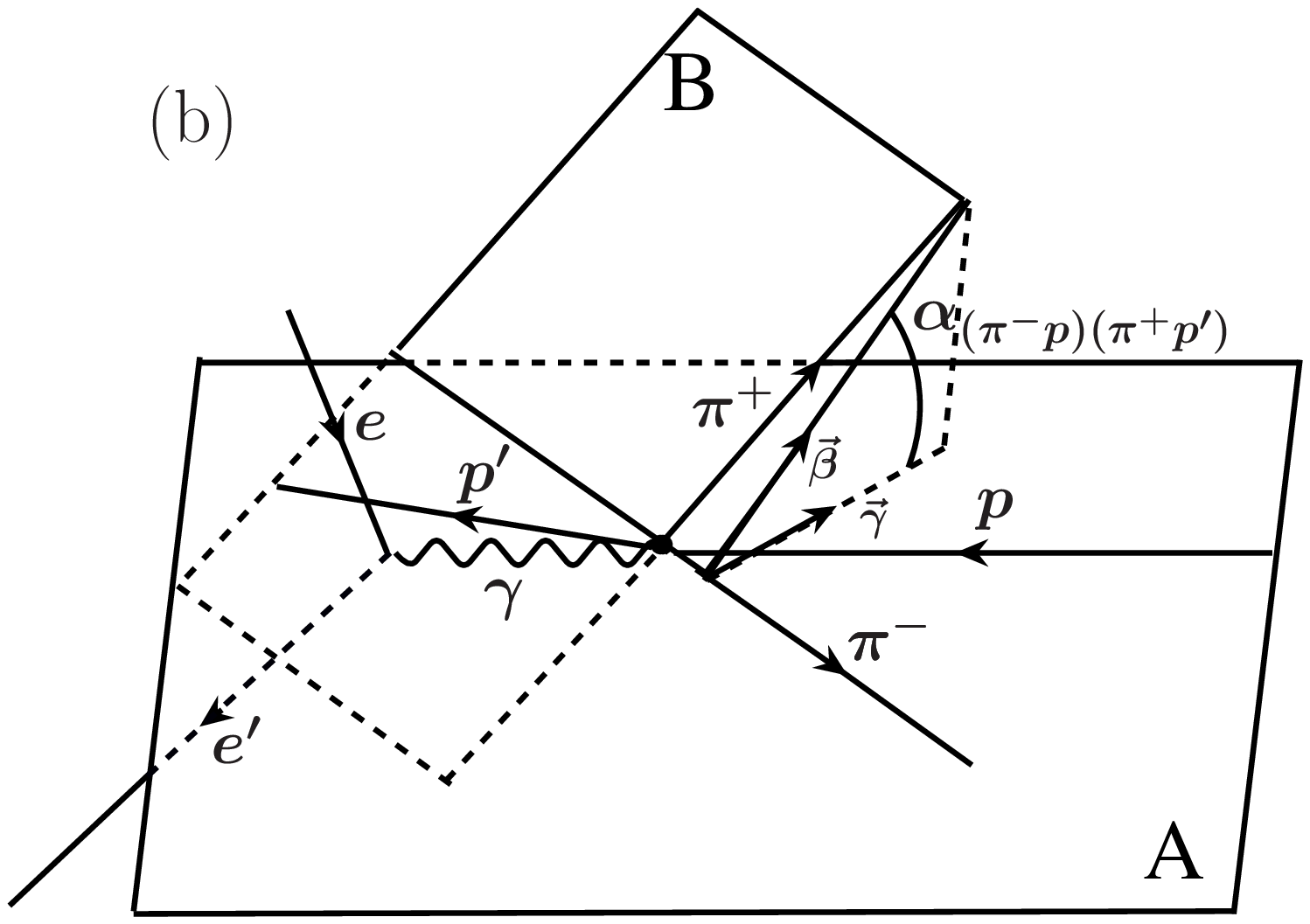}
\caption{\small Kinematic variables for the description  of
$e p \rightarrow e' p' \pi^{+} \pi^{-}$ in the CM frame of the final-state 
hadrons corresponding to the first assignment presented
in Section~\ref{kinxsect}. Panel (a) shows the 
$\pi^{-}$ spherical angles $\theta_{\pi^{-}}$ and $\varphi_{\pi^{-}}$. Panel (b)
shows the angle $\alpha_{[p\pi^{-}][p'\pi^{+}]}$  between the two planes: one of them (plane A)
is defined by the 3-momenta of the initial proton and the final $\pi^{-}$,
the other (plane B) is defined by the 3-momenta of the two others final
hadrons $\pi^{+}$ and proton. The unit vectors 
$\overline{\gamma}$ and $\overline{\beta}$ are normal
to the $\pi^-$ three-momentum in the planes A and B, respectively.} \label{kinematic}
\end{center}
\end{figure}

\begin{figure*}[ht]
\begin{center}
\hspace{-2cm}
\vspace{0.5cm}
\epsfig{file=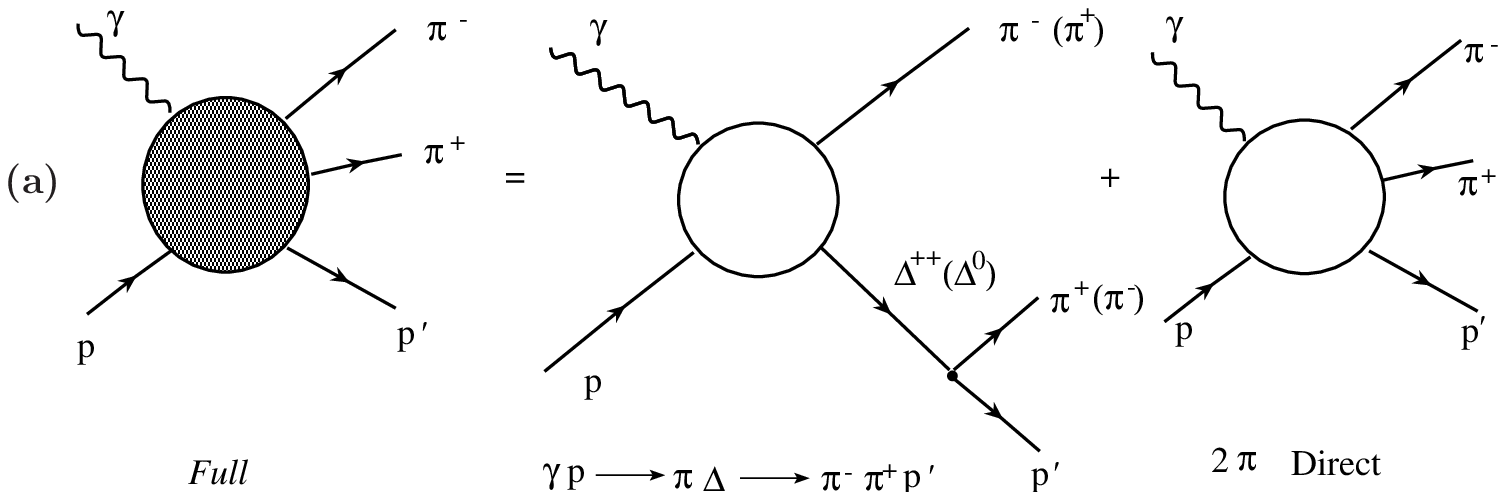,width=12cm}
\end{center}
\begin{center}
\hspace{-3cm}
\vspace{0.5cm}
\epsfig{file=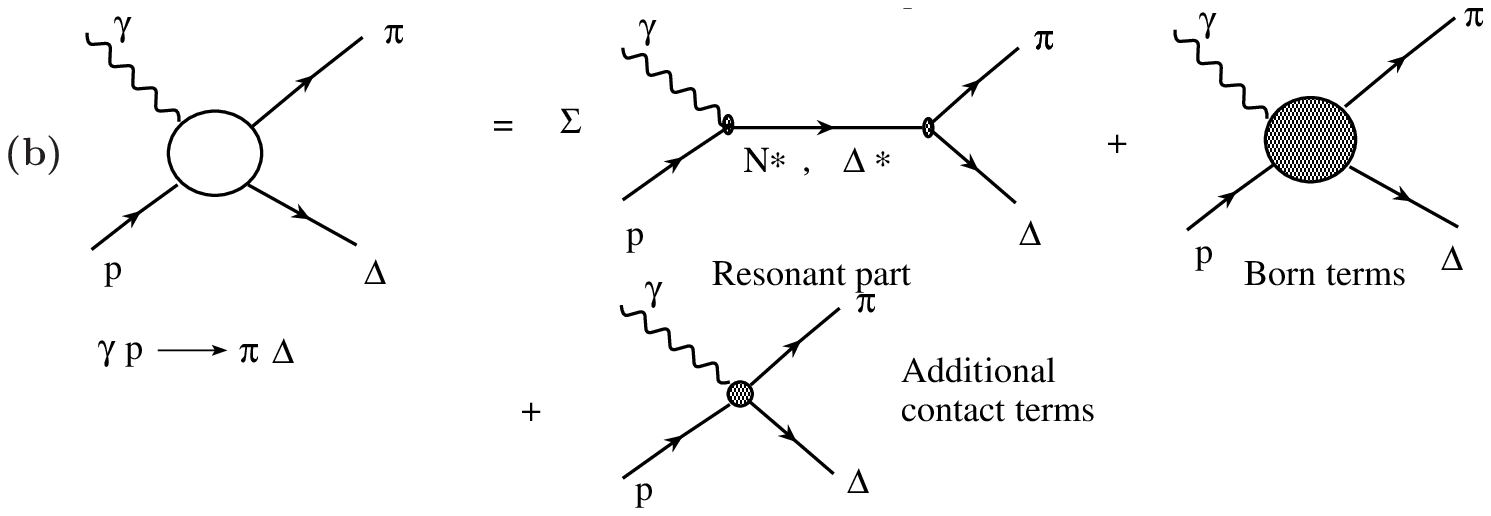,width=12cm}
\end{center}
\begin{center}
\hspace{-3.cm}
\vspace{-1.cm}
\epsfig{file=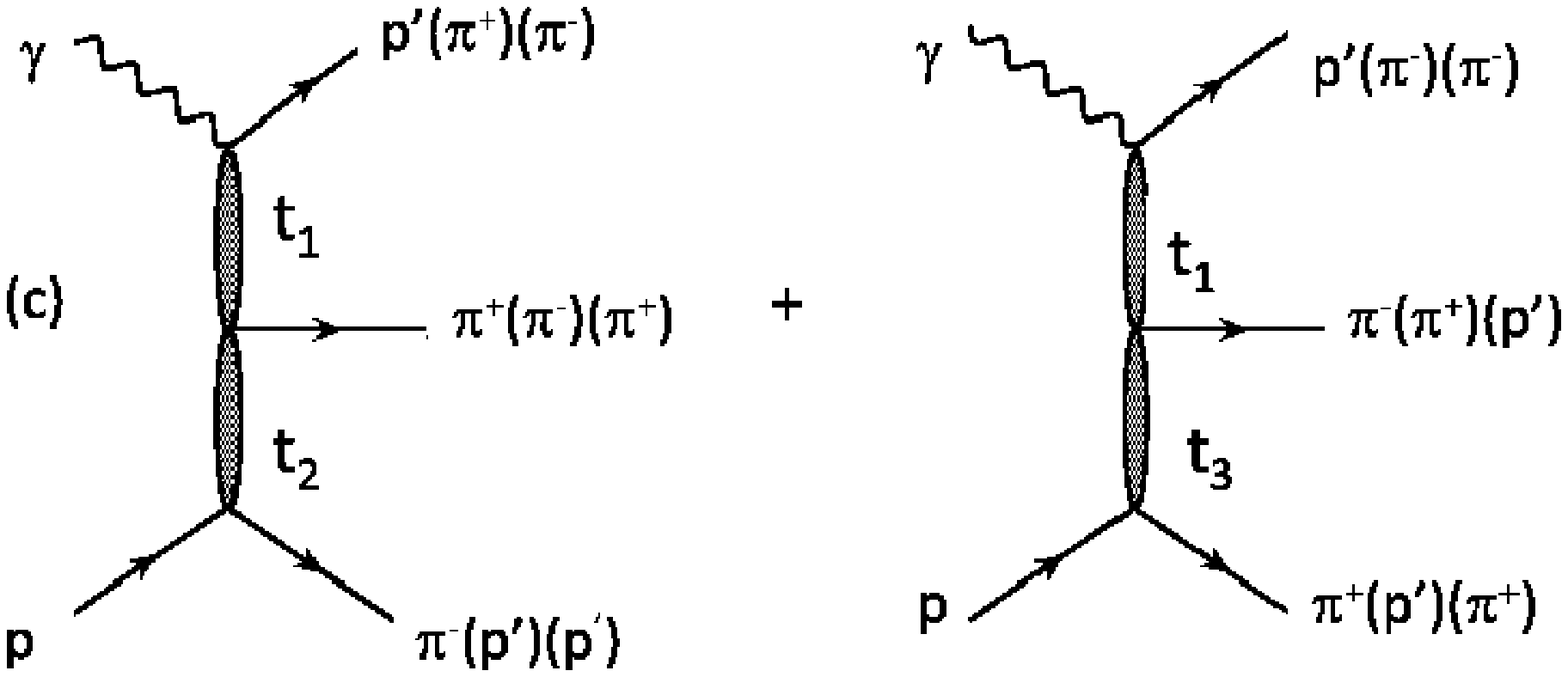,width=10.cm}
\end{center}
\begin{center}
\hspace{-3.cm}
\vspace{-0.5cm}
\epsfig{file=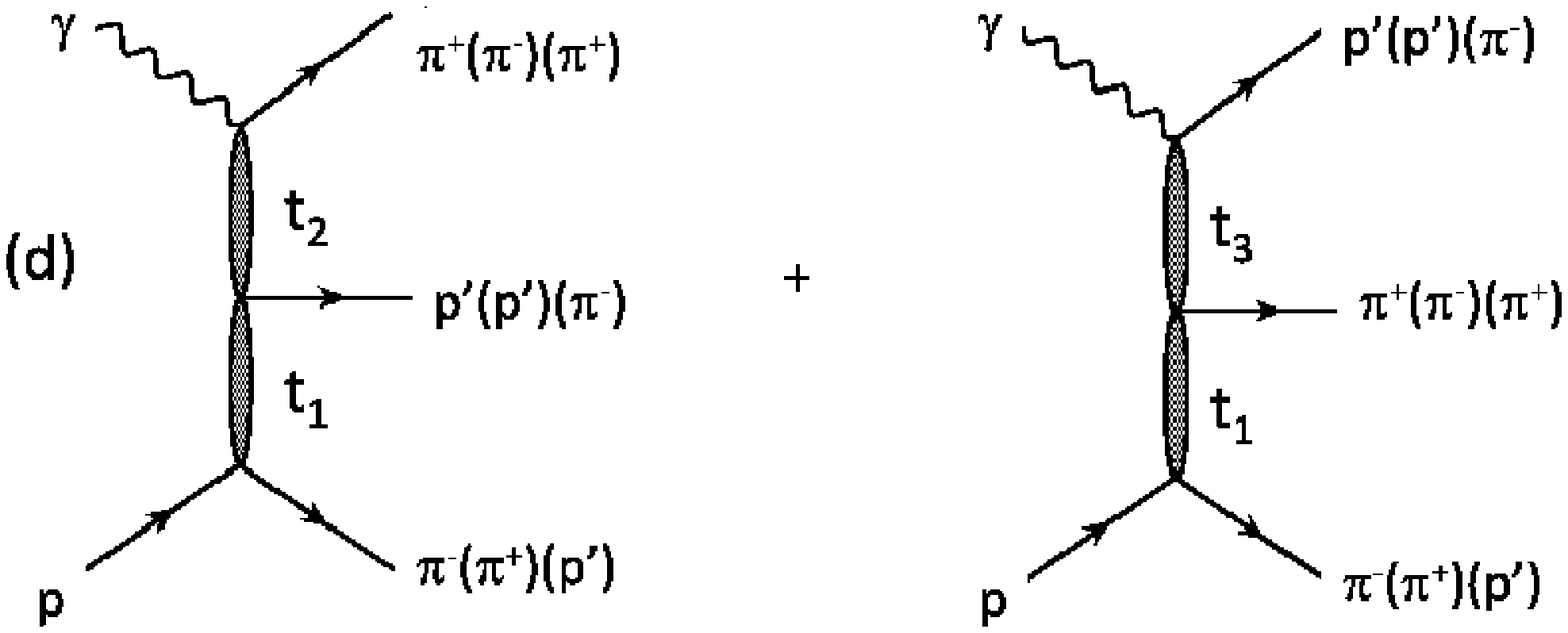,width=10.cm}
\end{center}
\caption{\small Reaction mechanisms of the JM model \cite{Mo09} that contribute to $\pi^{+}\pi^{-} p$ 
electroproduction in the kinematic region covered by recent CLAS
measurements \cite{Fe09} with W $<$ 1.6 GeV and 0.25 GeV$^{2}$ $<$ $Q^{2}$ $<$
0.6 GeV$^{2}$: 3-body mechanisms (a), $\pi \Delta$ isobar channels (b), and 
direct 2$\pi$ production 
mechanisms that correspond to different assignments for the 
final state hadrons (c,d).
 The $t_{i}$ ($i$=1,..3) stand for the squared transferred momenta in 
 the exchange processes by
 unspecified particle(s), as described in \cite{Mo09} and shown by blobs in the (c)
 and (d) panels.}
\label{mech_low_06}
\end{figure*}

\section{Phenomenological model JM for evaluation of $\gamma_{v}pN^*$ electrocouplings}
\label{genjm}
The phenomenological meson-baryon model JM was developed to describe 
$\pi^+\pi^-p$ electroproduction off protons \cite{Mo09,Mo07,Sh07,Mo06,Az05,Mo06-1} 
with 
the primary objective of determining the resonance  $\gamma_{v}pN^*$ electrocouplings 
and the $\pi\Delta$ and $\rho p$ 
partial hadronic decay widths from a combined fit to all
measured observables. In our current analysis of the CLAS $\pi^+\pi^-p$ electroproduction data \cite{Fe09} the JM model was used to separate the resonant and non-resonant contributions 
to differential cross sections and to access the electrocouplings and $\pi \Delta$, 
$\rho p$ decay widths of the $P_{11}(1440)$ and $D_{13}(1520)$ resonances.
Here we briefly discuss the basic ingredients of 
the JM model that
are relevant to the objectives of this paper.

\subsection{Kinematics and cross sections}
\label{kinxsect}
At a given invariant mass $W$ and photon virtuality $Q^2$,
the $\gamma_{v} p \rightarrow \pi^+\pi^-p$ reaction
can be fully described as a five-fold differential cross section 
$d^5\sigma/d^5\tau$,
where $d^5\tau$ is the phase-space volume of the
five independent variables in the center-of-mass (CM) system of the final
$\pi^{+} \pi^{-} p$ state. There are
many possible choices \cite{Byc} of the
five independent variables.
After defining $M_{\pi^+p}$, $M_{\pi^-p}$, and
$M_{\pi^+\pi^-}$ as invariant mass variables of
 the three possible two-particle pairs in the $\pi^{+}\pi^{-}p$ system,
we adopt the following three assignments:

\begin{enumerate}
\item $d^5\tau_1=dM_{p\pi^+}dM_{\pi^+\pi^-}d\Omega_{\pi^-}
d\alpha_{[p'\pi^{+}][p\pi^{-}]}$, where
$\Omega_{\pi^-}$ ($\theta_{\pi^{-}}$, $\varphi_{\pi^{-}}$) are
the final $\pi^-$
spherical angles with respect to the direction of the virtual photon, and
$\alpha_{[p' \pi^{+}][p\pi^{-}]}$ is the angle
between Plane B defined by the momenta of
the final $p'\pi^{+}$ pair and Plane A defined by the momenta of the initial proton and the final $\pi^{-}$ ;
\item $d^5\tau_2=dM_{p\pi^+}dM_{\pi^+\pi^-}d\Omega_{p'}d\alpha_{[\pi^{+}\pi^{-}][p'p]}$,
where 
 $\Omega_{p'}$ ($\theta_{p'}$, $\varphi_{p'}$) are the final proton
spherical angles with respect to the direction of the virtual photon, and
 $\alpha_{[\pi^{+}\pi^{-}][p'p]}$ is the angle
between Plane B$^{'}$ defined by the momenta of the 
$\pi^{+}\pi^{-}$ pair and Plane A  defined by the momenta of the
initial and final protons ;
\item $d^5\tau_3=dM_{p\pi^+}dM_{p\pi^-}d\Omega_{\pi^+}
d\alpha_{[p'\pi^{-}][p\pi^{+}]}$, where 
$\Omega_{\pi^+}$ ($\theta_{\pi^{+}}$, $\varphi_{\pi^{+}}$)
are 
the final $\pi^+$
spherical angles with respect to the direction of virtual photon,
and
$\alpha_{[p'\pi^{-}][p\pi^{+}]}$ is the angle 
between  Plane B$^{''}$ defined by the momenta of
the final $p'\pi^{-}$ pair and Plane A$^{''}$ defined by the momenta of the initial proton and the $\pi^{+}$.
\end{enumerate}
All frame-dependent variables are defined in the final hadron center-of-mass (CM)
frame.

The emission angles for the final state particles in the case of the first assignment 
are shown in  Fig.~\ref{kinematic}.
This choice is 
most suitable for describing  $\pi^+\pi^-p$ electroproduction through the 
$\pi^{-}\Delta^{++}$ intermediate state, which is the dominant contributor
of all isobar channels in  the kinematic region covered by the data \cite{Fe09}.
 For the other assignments the emission
angles of the final hadrons are analogous to the ones given in Fig.~\ref{kinematic}.
The relations between the  momenta of the final-state hadrons and the five
variables of the first assignment can be found in the Ref.~\cite{Fe09}.

The $\pi^+\pi^-p$ electroproduction data have been collected in the 
bins of a seven
dimensional space. As mentioned above, five variables are needed to fully
describe the final hadron kinematics, while to describe 
the initial state kinematics two others variables 
$W$ and $Q^2$ are required.  The huge number of seven dimensional 
bins over the reaction phase space ($\approx$ 100,000 bins) 
does not allow us to use the
correlated multi-fold differential cross sections in the analysis of
electroproduction processes, where the statistics decrease drastically with the photon 
virtualities $Q^2$. More than half of the five-dimensional phase-space bins 
of the final hadrons  is not populated due to statistical limitations. 
This is a serious obstacle 
for any analysis method that employs information on the behavior of multi-fold differential cross sections. 
We therefore  use the following one-fold differential cross sections in each bin of 
$W$ and $Q^2$ covered by the measurements:
\begin{itemize}
\item invariant mass distributions for three pairs of the final particles 
$d\sigma/dM_{\pi^{+}\pi^{-}} $, $d\sigma/dM_{\pi^+ p}$, and 
$d\sigma/dM_{\pi^- p}$;
\item angular distributions for spherical angles of the three final particles $d\sigma/d(-cos\theta_{\pi^-})$,
$d\sigma/d(-cos\theta_{\pi^+})$, and $d\sigma/d(-cos\theta_{p'})$ 
in the CM frame;
\item angular distributions for the three $\alpha$-angles described above and determined in the CM frame 
$d\sigma/d\alpha_{[p'\pi^+][p\pi^{-}]}$,  $d\sigma/d\alpha_{[p'\pi^-][p\pi^{+}]}$, and
 $d\sigma/d\alpha_{[\pi^+\pi^-][p p']}$.
\end{itemize} 
The one-fold differential cross sections were obtained from integrating the 
five-fold differential 
cross sections over the relevant four others kinematic 
variables of $d^{5}\tau_{i}$. All details related to the evaluation of the $\pi^+\pi^-p$ one-fold differential cross sections we are using for the extraction of
resonance parameters can be found in Ref.~\cite{Fe09}.

\subsection{Relevant electroproduction mechanisms}
\label{pipipmech}

The major part of the $\pi^+\pi^-p$ electroproduction off 
protons  at $W$ $<$ 1.6 GeV is due to 
contributions from the two $\pi \Delta$ isobar channels $\pi^- \Delta^{++}$ and $\pi^+ \Delta^0$.
The $\Delta^{++}$(1232) resonance
is clearly seen in all $\pi^{+} p$ mass distributions for W $>$ 1.4
GeV, while contributions from the $\pi^+ \Delta^0$ isobar channel are needed to better describe
the data in the low mass regions of the $\pi^- p$ mass distributions. 
The observed in the data \cite{Fe09} strength of the $\pi^- \Delta^{++}$ isobar 
channel is approximately nine times larger than that of $\pi^+ \Delta^0$ \cite{Mo09} 
due to isospin invariance. 
The contributions from all other isobar channels $p\,\rho$,
$\pi^{+}D_{13}^{0}(1520)$, $\pi^{+}F_{15}^{0}(1685)$, and
$\pi^{-}P_{33}^{++}(1640)$, which are incorporated into the JM-model \cite{Sh07,Mo06,Az05,Mo06-1} in order to describe the data at $W$ $>$ 1.6 GeV, 
are negligible in the kinematic region covered in this analysis, and are not included 
in this work.

The production amplitudes relevant for our
analysis of $\gamma_{v} p \rightarrow \pi^+\pi^- p$ are illustrated in Fig.~\ref{mech_low_06}. 
They consist of the $\pi^-
\Delta^{++}$ and $\pi^{+} \Delta^{0}$ isobar channels and direct double pion
production mechanisms. The production amplitudes for $\pi \Delta$ intermediate
states (Fig.~\ref{mech_low_06}b) consist of the resonant contributions 
$\gamma_{v} N \rightarrow N^*,\Delta^* \rightarrow \pi\Delta$ and non-resonant terms. 
All resonances listed in Table~\ref{nstlist} are
included in the JM model. However, in the kinematic area 
covered in our measurements, only the 
$P_{11}(1440)$,
$D_{13}(1520)$, and $S_{11}(1535)$ nucleon resonances have strength that is sufficient to manifest themselves in the 
one-fold differential
cross-sections. 
Non-resonant amplitudes, depicted in the diagrams in  Fig.~\ref{mech_low_06}b are computed from the
 well-established Born terms 
 presented in Appendix A of Ref.~\cite{Mo09}. The additional 
contact terms are implemented in the full $\pi \Delta$ production amplitudes. 
They describe effectively the  contributions from mechanisms other
than the Born terms to $\pi \Delta$ production and the part of the  
$\pi \Delta$ final-state interactions (FSIs) that are not included in
the JM model's absorptive approximation for FSIs \cite{Mo09,Ri00}.
The Lorentz structure of these additional contact terms is determined by
superposition of the two second-rank Lorentz tensors
\begin{eqnarray}
\label{newstr}
\gamma^{\mu}p^{\pi}_{\nu},\nonumber  \\
p_{c}^{\delta}\gamma_{\delta}g_{\mu\nu},
\end{eqnarray}
where 
$p_{c}=(2p^{\pi}_{\nu}-q_{\gamma})$ is the difference of the final pion 
four-momentum $p^{\pi}_{\nu}$ and 
the momentum transferred, $q_{\gamma}-p^{\pi}_{\nu}$. 
A parametrization of the additional 
contact-term
amplitudes in the $\pi \Delta$ channels may be found in Appendix B of Ref.~\cite{Mo09}.

\begin{table}
\begin{center}
\begin{tabular}{|c|c|c|c|c|c|}
\hline
$N^{*}$ states & Mass,\ & Total   &
Branching  & Branching  & $N^{*}$ electro  \\
incorporated &  (GeV)  & decay    &
ratio    & ratio  & coupling  \\
into the  &   & width  & ${\pi\Delta}$, \% & ${\rho p}$, \% & variation  \\
data fit&  & $\Gamma_{tot}$,  &  &  &  in the fit \\
        &  &        (GeV)       &  &  &             \\ 
\hline
$P_{11}(1440)$ & var & var & var & var & \cite{Az051} var \\
$D_{13}(1520)$ & var & var & var & var & \cite{Az051} var \\
$S_{11}(1535)$ & var & var & var & var & \cite{Az09} fix \\
$S_{31}(1620)$ & 1.62 & 0.16 & 60 & 16 & \cite{Az05,Mo06} var\\
$S_{11}(1650)$ & 1.65 & 0.15 & 2 & 3 & \cite{Bu03} var \\
$F_{15}(1680)$ & 1.68 & 0.12 & 12 & 5.5 & \cite{Az05,Mo06} var \\
$D_{13}(1700)$ & 1.74 & 0.19 & 53 & 45 & \cite{Az05,Mo06} fix \\
$D_{33}(1700)$ & 1.70 & 0.26 & 89 & 2 & \cite{Az05,Mo06} fix  \\
\hline
\end{tabular}
\caption{\label{nstlist} List of resonances invoked in the $\pi^+\pi^-p$ 
fit and their parameters: total decay widths $\Gamma_{tot}$ 
and branching fractions (BF) to $\pi
\Delta$ and $\rho p$ final states.
The quoted values for the hadronic parameters are taken from fits to 
the earlier CLAS $\pi^+\pi^-p$ data \cite{Ri03} 
using the 2005
version of the JM model \cite{Az05,Mo06}.
The quantities labeled as $var$ correspond to the variable parameters fit
to the CLAS $\pi^+\pi^-p$ data \cite{Fe09} within the framework of the 
current JM model version \cite{Mo09} 
employing the unitarized BW ansatz of Section~\ref{unitbw} for the resonant
contributions. Start values for the resonance electrocouplings are taken from
the references listed in the last column and extrapolated to the $Q^2$ area
covered by the CLAS experiment \cite{Fe09}.}
\end{center}
\end{table}

All isobar channels combined account for 70\% to 90\% of the charged double-pion 
fully integrated
cross sections in the kinematic region covered by data \cite{Fe09}.
The remaining part of the cross sections comes from 
direct charged  pion (2$\pi$) production mechanisms, in which the $\pi^{+} \pi^{-} p$ final state 
 is created 
without the formation
of unstable hadrons in the intermediate states. The presence of these mechanisms 
is required by the unitarity of 
the three-body
$\pi^+\pi^-p$ production amplitudes \cite{Ait78}. 
Their manifestation in $\pi^{+} \pi^{-} p$
electroproduction was observed for the first time in our previous analyses 
of CLAS data \cite{Az05,Mo09}. 
The dynamics of these
processes was unknown and has been established from the CLAS data analysis within the framework of the 
JM model. The direct  
2$\pi$-production mechanisms incorporated in the JM model are depicted 
in Fig.~\ref{mech_low_06}(c,d). They represent 
two subsequent
exchanges of unspecified particles, parametrized by propagators that depend exponentially on the running
four-momenta squared. Each set of diagrams in Fig.~\ref{mech_low_06} 
corresponds to various assignments of the final-state hadrons, resulting in 
different four-momenta
squared running over propagators in the exchange amplitudes. The JM model extends for the first 
time the description of 
$\pi^+\pi^-p$ electroproduction beyond the approximation of superimposed 
isobar channels by incorporating direct 2$\pi$-production. 
Explicit expressions for the above-mentioned direct 2$\pi$-production
amplitudes can be found in Appendices A-C of Ref.~\cite{Mo09}.

The relationships between $\pi^+\pi^-p$ electroproduction cross sections 
and the three-body production amplitudes employed in the JM model are given 
in Appendix D of Ref.~\cite{Mo09}. This information is required in order to compare the amplitudes of the JM model with the results from any other study of $\pi^+\pi^-p$ electroproduction amplitudes. 

\subsection{Breit-Wigner parametrization of resonant amplitudes}
\label{regbw}
We start from a non-unitarized relativistic Breit-Wigner (BW) ansatz to 
describe the resonant contribution $\langle \lambda_{f} \left| T_{r} \right| \lambda_{\gamma}\lambda_{p}
\rangle$ in the helicity representation:
\begin{eqnarray}
\label{regbwamp}
\langle \lambda_{f} \left| T_{r} \right| \lambda_{\gamma}\lambda_{p}
\rangle =
\sum_{N^{*}}\frac{\langle  \lambda_{f} \left| T_{dec} \right|\lambda_{R} \rangle 
\langle \lambda_{R} \left| T_{em} \right| \lambda_{\gamma}\lambda_{p} \rangle}
{M_{r}^{2}-W^{2}-i\Gamma_{r}(W)M_{r}} ,
\end{eqnarray}
where  \(M_{r}\) and \(\Gamma_{r}\) are the resonance mass and 
energy-dependent total width, respectively. The matrix elements
\(\langle \lambda_{R} \left| T_{em} \right|\lambda_{\gamma}\lambda_{p}\rangle\) and     \(\langle  \lambda_{f} \left| T_{dec} \right| \lambda_{R} \rangle\)
are the electromagnetic production and hadronic decay
amplitudes of the \(N^{*}\) with helicity
\(\lambda_{R}=\lambda_{\gamma}-\lambda_{p}\), in which  $\lambda_{\gamma}$ and
$\lambda_{p}$ stand for the helicities of the photon and proton in the initial
state, and $\lambda_{f}$
represents the helicities of final-state hadrons in the $N^*$ decays.

The  hadronic decay
amplitudes $\langle \lambda_{f} \vert T_{dec} \vert 
\lambda_{R} \rangle$ are related to the $\Gamma_{\lambda_{f}}(W)$ partial hadronic decay widths of the $N^{*}$ 
to $\pi \Delta$ or $\rho p$
final states $f$ of helicity $\lambda_{f}$ by: 
\begin{equation}
\label{decaywidth1}
\langle \lambda_{f} \vert T_{dec} \vert \lambda_{R} \rangle=\langle \lambda_{f} \vert T^{J_{r}}_{dec} \vert 
\lambda_{R} \rangle d^{J_{r}}_{\mu\nu}(\cos\theta^{*})e^{i\mu\phi^{*}}, \nonumber \\
\end{equation}
with $\mu=\lambda_{R}$  and 
$\nu=-\lambda_{\Delta}$   for the  $\pi \Delta$    intermediate state  \\
and $\nu=\lambda_{p'}-\lambda_{\rho}$    for the  $\rho$p$^{'}$ 
intermediate state, and 
\begin{equation}
\label{decaywidth5} 
\langle \lambda_{f} \vert T^{J_{r}}_{dec} \vert \lambda_{R}
\rangle=\frac{2\sqrt{2\pi}\sqrt{2J_{r}+1}M_{r}\sqrt{\Gamma_{\lambda_{f}}}}{\sqrt{\langle
p^{r}_{i} \rangle}}\sqrt{\frac{\langle p^{r}_{i} \rangle}{\langle p_{i} \rangle}}  . 
\end{equation}
The means $\langle p^{r}_{i} \rangle$ and $\langle p_{i} \rangle$ are the 
magnitudes of the 
three-momenta  of the final $\pi$ for the   
$N^* \rightarrow \pi \Delta$ decay ($i$=1) or of the final proton  
for the $N^* \rightarrow \rho p$ decay ($i$=2), evaluated at $W=M_{r}$ and 
at the running $W$, respectively, and  
averaged over the running mass of the unstable hadron in the
intermediate state. The variables $\theta^{*}$, $\phi^{*}$ are the CM  polar and azimuthal 
angles for the final $\pi$ ($N^* \rightarrow \pi \Delta$ decay) or the final
proton ($N^* \rightarrow \rho p$ decay). The symbol $J_{r}$ stands for the $N^*$ spin. 
This relationship between the $N^*$ hadronic decay
amplitudes and the partial decay widths was derived for the 
$N^* \rightarrow \pi \Delta$ decay in our previous article \cite{Ri00}. All details on the parametrization of the resonance 
hadronic decay amplitudes in the JM model  
can be found in Appendix~A of this paper.

Energy dependencies of the partial hadronic decay widths are described
under the assumption that centrifugal barrier-penetration factors are the major contributors to the off-shell behavior of resonance hadronic decay amplitudes \cite{Ri00,Bla52}:  
\begin{gather}
\sqrt{\Gamma_{LS}} = \sqrt{\Gamma^{r}_{LS}}\cdot
\left[ \frac{M_{r}}{W}
\frac{(J_{l}^{2}(p^{r}R) + N_{l}^{2}(p^{r}R))}
{(J_{l}^{2}(pR) + N_{l}^{2}(pR))} \right]^{1/2},
\label{ww}
\end{gather}
where \(J_{l}\) and \(N_{l}\) are the Bessel and Neumann functions. The  factor  in
square  brackets represent the ratio of barrier penetration factors for the final 
meson with orbital angular momentum \(l\) evaluated at $W=M_{r}$ and 
at the running $W$.  
The variable R represents the interaction radius whose value was set to 1 fm, 
and $\sqrt{\Gamma^{r}_{LS}}$ stands for the decay
amplitude estimated at the resonant point. The partial decay 
amplitudes $\sqrt{\Gamma_{LS}}$ of Eq. (\ref{ww}) 
are transformed from the $LS$ to the helicity representation
$\sqrt{\Gamma_{\lambda_{f}}}$
(see the Appendix A) 
and used to compute the $N^*$ hadronic decay amplitudes via 
Eq. (\ref{decaywidth5}).

In the  JM model, the \(N^{*}\) total decay width \(\Gamma_{r}(W)\) 
in Eq.~(\ref{regbwamp}) is 
evaluated as a sum  
over all partial decay widths. In this way we ensure unitarity of 
the resonant amplitude for a single resonance
contribution. The unitarization procedure in the actual case of many contributing 
resonances will be
discussed in Section \ref{unitbw}.

The resonance electroexcitation amplitudes
\(\langle \lambda_{R} \left| T_{em}
\right| \lambda_{\gamma}\lambda_{p} \rangle\) in Eq.~(\ref{regbwamp}) are related to the 
$\gamma_{v}NN^*$ electrocouplings \(A_{1/2}\),
\(A_{3/2}\),
and \(S_{1/2}\) for nucleons. The definition of these electrocouplings in the JM model is consistent with
the Review of Particle Physics (RPP) \cite{PDG10} relation between the \(A_{1/2}\),
\(A_{3/2}\) electrocouplings and the $N^*$ electromagnetic decay
width $\Gamma_{\gamma}$: 
\begin{equation}
\label{a12a32width1} 
\Gamma_{\gamma}=
\frac{q^{2}_{\gamma, r}}
{\pi}\frac{2M_{N}}{(2J_{r}+1)M_{r}} \left[ \left | A_{1/2} \right |^{2}+\left |
A_{3/2} \right|^{2} \right]  ,
\end{equation} 
where $q_{\gamma, r}$ is the three-momentum modulus of the photon at $W=M_{r}$ 
in the CM frame. 
The  
transition amplitudes 
$\langle \lambda_{R} \left| T_{em} \right| \lambda_{\gamma}\lambda_{p} \rangle$ are related to the $\gamma_{v}NN^*$ $A_{1/2}$, $A_{3/2}$, and $S_{1/2}$ electrocouplings by imposing the requirement that the 
BW parametrization \cite{LS71} of the resonant cross section 
for a single contributing resonance should be reproduced:
\begin{gather}
\label{secbw} 
\sigma(W)=\frac{\pi}{q_{\gamma}^2}(2J_{r}+1) 
\frac{M^{2}_{r}\Gamma_{i}(W)\Gamma_{\gamma}}
{(M^{2}_{r}-W^{2})^{2}-M^{2}_{r}\Gamma^{2}_{r}(W)}\frac{q_{\gamma}}{K}. 
\end{gather}
Here the photon equivalent energy 
$K=\frac{W^2-M_{N}^{2}}{2W}$, $q_{\gamma}$ is the absolute value of the initial photon three-momentum of
virtuality $Q^2$ $>$ 0  $q_{\gamma}=\sqrt{Q^2+E_{\gamma}^2}$ and the energy $E_{\gamma}$
in the CM frame at the running $W$
\begin{eqnarray}
\label{ccc}
E_{\gamma}=\frac{W^2-Q^2-M_{N}^2}{2W}. 
\end{eqnarray}
The $q_{\gamma, r}$ value in Eq. (\ref{a12a32width1}) can be computed 
from
Eq. (\ref{ccc}) at W=$M_{r}$.  
$\Gamma_{i}(W)$ is the energy-dependent hadronic decay width to the final state $\pi\Delta$ ($i$=1) 
or $\rho p$ ($i$=2). The factor $\frac{q_{\gamma}}{K}$ in Eq. (\ref{secbw}) is equal to
unity at the photon point. It accounts for the choice \cite{Fe09} of the virtual
photon flux in the evaluation of the virtual photon cross sections. 
In this way we obtain the following relationship between the transition amplitudes 
$\langle \lambda_{R} \left| T_{em} \right| \lambda_{\gamma}\lambda_{p} \rangle$ and the 
$\gamma_{v}NN^*$ electrocouplings:
\begin{eqnarray}
\label{a12a32sec}
\langle \lambda_{R} \left| T_{em} \right| \lambda_{\gamma}\lambda_{p} \rangle  
=\frac{W}{M_{r}}\sqrt{\frac{8M_{N}M_{r}q_{\gamma_{r}}}{4\pi\alpha}}
\sqrt{\frac{q_{\gamma_{r}}}{q_{\gamma}}}A_{1/2,3/2}(Q^{2}), \nonumber \\
\text{with}\left| \lambda_{\gamma}-\lambda_{p} \right| = \frac{1}{2},\, \frac{3}{2} \,\, 
\text{ for transverse photons, and}\nonumber\\
\langle \lambda_{R} \left| T_{em} \right| \lambda_{\gamma}\lambda_{p} \rangle 
=\frac{W}{M_{r}}\sqrt{\frac{16M_{N}M_{r}q_{\gamma_{r}}}{4\pi\alpha}}
\sqrt{\frac{q_{\gamma_{r}}}{q_{\gamma}}}\, S_{1/2}(Q^{2}) , 
\nonumber \\
 \text{for\, longitudinal\, photons.}\nonumber\\
\end{eqnarray}
The factor
$4\pi\alpha$ in Eqs. (\ref{a12a32sec}) reflects the particular 
relationship between the JM model amplitudes and cross sections \cite{Mo09}, when
the absolute value of the electron charge is factorized out of the production
amplitudes.

\subsection{Unitarization of the full resonant amplitudes}
\label{unitbw}
\begin{figure}[t]
\includegraphics[width=9cm]{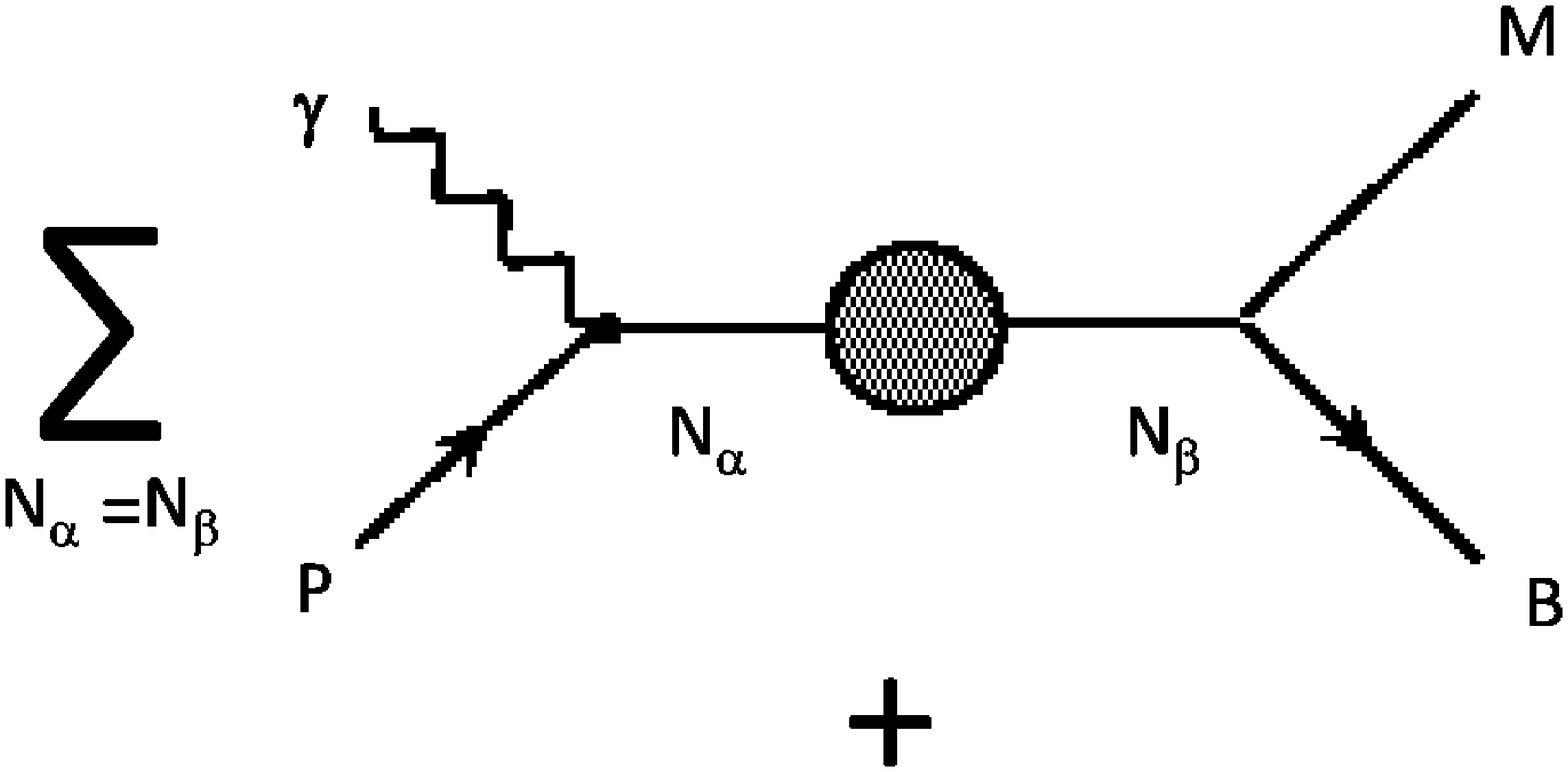}
\includegraphics[width=9cm]{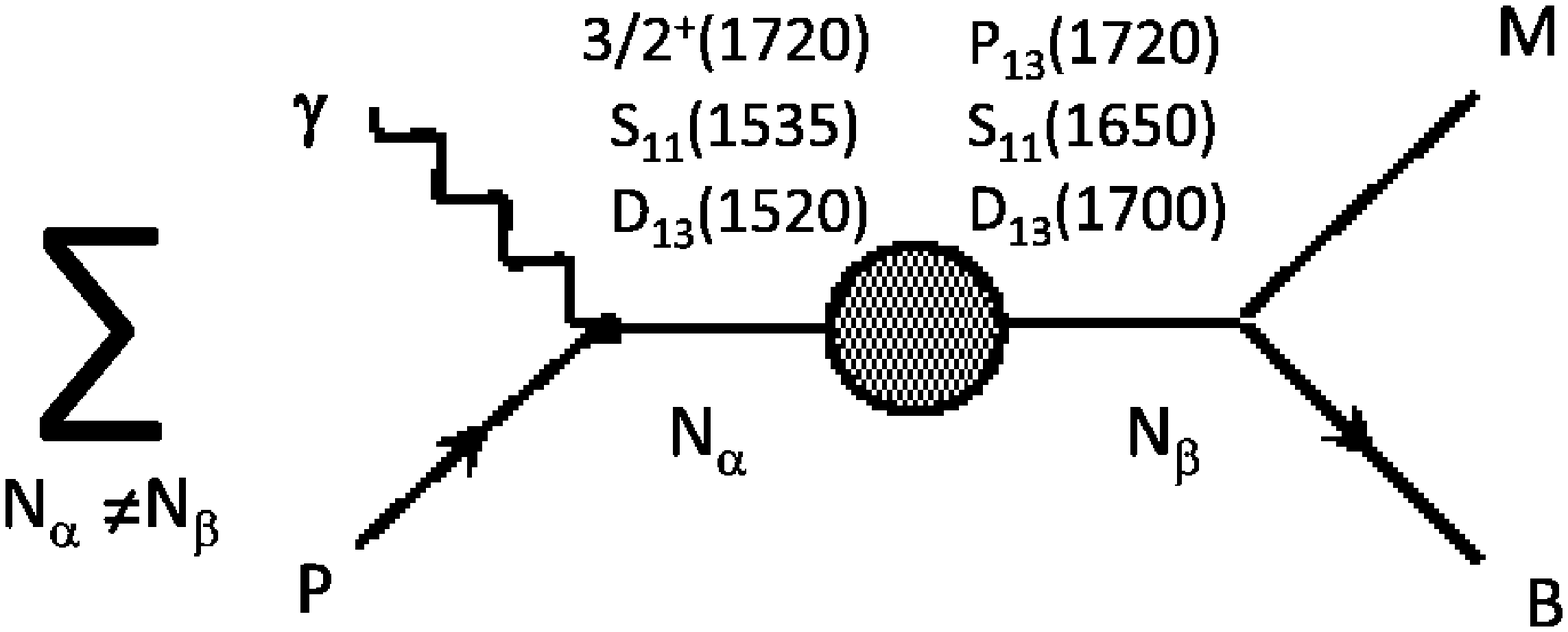}
\caption{\small Diagrams showing the unitarized BW ansatz of the JM model, which 
incorporates 
transitions
via the dressed resonance propagator between states listed in 
Table \ref{nstlist} for the same
$N_{\alpha}$=$N_{\beta}$ (top) and different  
$N_{\alpha} \ne N_{\beta}$ resonances as allowed by the conservation 
laws of the strong interaction (bottom).}
\label{fullresamp}
\end{figure}

The BW ansatz, described  in Section~\ref{regbw}, provides unitary contributions
from any individual $N^*$ state to the resonant part of the electroproduction
amplitudes. 
However, the full resonant amplitude, which represents a superposition of contributions from all $N^*$ states,
is not unitary. Figure \ref{fullresamp}  schematizes the processes that contribute to the 
resonant propagator dressing. They consist of transitions between the same and different $N^*$ states mediated by the strong interaction. The regular BW ansatz
incorporates only transitions between the same $N^*$ states shown 
by the blob in the top panel of  Fig.~\ref{fullresamp}. The unitarization is achieved via inclusion of both diagrams
in Fig. 3, allowing us to take into account 
transitions between
both the same and 
different $N^*$ states (bottom panel in Fig.~\ref{fullresamp}) in the dressed
resonant propagators. We replaced the usual BW ansatz by its 
unitarized extension originally
proposed in Ref.~\cite{Ait72}. This ansatz was adjusted for the description of 
the BW resonance 
electroproduction amplitudes of Eq. (\ref{regbwamp}) employed in the JM model.  
The full resonant amplitude of the unitarized extension of the BW ansatz satisfies the unitarity condition~\cite{Ait72}.

Unitarized resonance amplitudes in the JM model incorporate  photo- and 
electro-excitation amplitudes of all relevant
resonances $\alpha$, all possible transitions between the initial $N^*$ state $\alpha$ and
the final $N^*$ state $\beta$, and hadronic decays of  $\beta$. Fully unitarized resonance amplitudes are determined by the sum  
of the products of the electromagnetic excitation amplitudes of the $\alpha$-th $N^*$ state 
$\langle \lambda_{\alpha} \left| T_{em} \right| \lambda_{\gamma}\lambda_{p} \rangle$, the hadronic decay 
amplitudes of the $\beta$-th $N^*$ state 
$\langle  \lambda_{f} \left| T_{dec} \right|\lambda_{\beta} \rangle$, 
the dressed
propagator $S_{\alpha \beta}$, and
\begin{gather}
\langle \lambda_{f} \left| T_{res} \right| \lambda_{\gamma}\lambda_{p}
\rangle =
\sum_{\alpha,\beta}\langle \lambda_{f} \left| T_{dec} \right|
\lambda_{\beta} \rangle 
 S_{\alpha \beta} 
\langle \lambda_{\alpha} \left| T_{em} \right| \lambda_{\gamma}\lambda_{p} \rangle.
\label{matel}
\end{gather}
 The sum incorporates all transitions between the $\alpha$-th initial and the $\beta$-th
final $N^*$ states that are allowed by quantum number conservation laws.
It runs over all repeated indices $\alpha$ and $\beta$ that label the $N^*$ states. 
As a consequence of angular momentum conservation 
$\lambda_{\alpha}$=$\lambda_{\beta}$, but these two helicities stand either for the same $N^*$ states
if $\alpha$=$\beta$ (diagonal transitions) or different $N^*$ states if $\alpha \ne \beta$ 
(off-diagonal transitions).  
The index $f$ represents the final-state helicities, either for $\pi \Delta$ or for $\rho p$.

The expression for the inverse dressed propagator $S^{-1}_{\alpha \beta}$ is 
obtained  in Ref.~\cite{Ait72} (see Eq.~(6) in Ref.~\cite{Ait72}). However, the parametrization 
of the $N^*$ propagator in Ref.~\cite{Ait72} and in JM  are different.
  Therefore, we change the inverse resonant propagator 
  $S^{-1}_{\alpha \beta}$ of  Ref.~\cite{Ait72}, so that 
it coincides with the single $N^*$ contribution employed in the JM model.   
In this way we obtain 

\begin{eqnarray}
\label{transel}
S^{-1}_{\alpha \beta}=M^{2}_{N^*}\delta{\alpha\beta}
-i(\sum_{k}\sqrt{\Gamma_{\alpha_{k}}\Gamma_{\beta_{k}}})
\sqrt{M_{N^*_{\alpha}}M_{N^*_{\beta}}} \\
-W^2\delta_{\alpha\beta} \nonumber,
\end{eqnarray}
where the index $k$ represents the partial $N^*$ hadronic decay widths to all possible 
final states, decomposed over 
$LS$ partial waves. For a single $N^*$ contribution,
 Eq.~(\ref{transel}) 
coincides with the inverse BW propagator in Eq.~(\ref{regbwamp}).

The resonant contribution at $W$ $<$ 1.6 GeV  incorporates 
the $N^*$ states listed in Table~\ref{nstlist} and  the $3/2^+(1720)$ candidate state
observed in the previous analysis of the CLAS $\pi^+\pi^-p$ electroproduction data \cite{Ri03}. The conservation laws in strong interactions allow transitions in dressed resonant propagators only
between the following pairs of $N^*$ states: $D_{13}(1520)$ and $D_{13}(1700)$,
 $S_{11}(1535)$ and $S_{11}(1650)$, and  
 $3/2^+(1720)$ candidate and $P_{13}(1720)$ (bottom panel in Fig.~\ref{fullresamp}). 
Therefore, the unitarized resonant amplitude represents the sum of
regular BW amplitudes over all other $N^*$ states with only diagonal transitions in the dressed
resonance propagators plus contributions from the
aforementioned pairs of $N^*$ states that have both diagonal and non-diagonal transitions in the resonant
propagators. The contributions  from each pair of $N^*$ states 
$\langle \lambda_{f} \left| T_{2res} \right| \lambda_{\gamma}\lambda_{p}
\rangle$ 
are determined by  Eq.~(\ref{matel}) with indices $\alpha$ and $\beta$ running from
1 to 2, resulting in 2x2 $S^{-1}_{\alpha \beta}$ matrices. 
After invertion of  $S^{-1}_{\alpha \beta}$ in Eq.~(\ref{transel}) and insertion  
into Eq.~(\ref{matel}),
we get the following expressions for the contribution from each of the three pairs of 
$N^*$ states listed in Fig.~\ref{fullresamp}: 

\begin{gather}
\langle \lambda_{f} \left| T_{2res} \right| \lambda_{\gamma}\lambda_{p}
\rangle 
=  \frac{1}{det[S^{-1}_{\alpha \beta}]} * \nonumber\\
\{\langle \lambda_{f} \left| T_{dec} \right|
\lambda_{1} \rangle 
(M^2_{2}-i\Gamma_{2}(W)M_{2}-W^2) 
\langle \lambda_{1} \left| T_{em} \right| \lambda_{\gamma}\lambda_{p} \rangle \nonumber \\
+\langle \lambda_{f} \left| T_{dec} \right|\lambda_{2} \rangle 
(M^2_{1}-i\Gamma_{1}(W)M_{1}-W^2)
\langle \lambda_{2} \left| T_{em} \right| \lambda_{\gamma}\lambda_{p} \rangle \nonumber \\
+i\langle \lambda_{f} \left| T_{dec} \right|
\lambda_{2} \rangle \sum_{k}\sqrt{\Gamma_{1_{k}}\Gamma_{2_{k}}} 
\sqrt{M_{N^*_{1}}M_{N^*_{2}}}\langle \lambda_{1} \left| T_{em} \right| \lambda_{\gamma}\lambda_{p} \rangle \nonumber\\ 
+i\langle \lambda_{f} \left| T_{dec} \right|
\lambda_{1} \rangle\sum_{k}\sqrt{\Gamma_{1_{k}}\Gamma_{2_{k}}} 
\sqrt{M_{N^*_{1}}M_{N^*_{2}}}\langle \lambda_{2} \left| T_{em} \right| \lambda_{\gamma}\lambda_{p}
\rangle \}  .
\label{amp2res}
\end{gather} 
The first two terms in Eq.~(\ref{amp2res}) correspond to 
the diagonal $S_{\alpha \beta}$ elements. They describe the processes depicted 
in  the top panel 
of  Fig.~\ref{fullresamp}. The two other terms correspond to transition between different $N^*$ states, 
shown in the bottom panel of Fig.~\ref{fullresamp}. The determinant of 
$S^{-1}_{\alpha \beta}$ can be computed from Eq. (\ref{transel}) as
\begin{gather}
det[S^{-1}_{\alpha
\beta}]=(M^2_{1}-i\Gamma_{1}(W)M_{1}-W^2)\cdot \nonumber\\
(M^2_{2}-i\Gamma_{2}(W)M_{2}-W^2)
+(\sum_{k}\sqrt{\Gamma_{1_{k}}\Gamma_{2_{k}}})^2M_{1}M_{2}.
\label{determ}
\end{gather} 
It follows from Eqs. (\ref{amp2res}) and (\ref{determ}), that if transitions between two different 
$N^*$ states in the dressed resonant propagators become impossible, then
\begin{equation}
\sum_{k}\sqrt{\Gamma_{1_{k}}\Gamma_{2_{k}}}=0,
\end{equation}
namely, all off-diagonal terms become zero, while the diagonal terms give rise to the amplitudes of the regular 
BW ansatz. When the transitions between the pairs of $N^*$ states are turned on, they not only add off-diagonal
terms to Eq.~(\ref{amp2res}), but they also change the effective propagator in the 
$\langle \lambda_{f} \left| T_{2res} \right| \lambda_{\gamma}\lambda_{p}\rangle$ amplitude with respect to the BW ansatz Eq.~(\ref{regbwamp}), determined by
 1/$det[S^{-1}_{\alpha \beta}]$. 

\section{The CLAS data fit}
\label{fit}

The resonance parameters obtained in this paper 
 are fit to the CLAS $\pi^+\pi^-p$ electroproduction 
differential cross sections  \cite{Fe09}. A realistic
evaluation of resonance electrocoupling uncertainties extracted in the fit
is an important objective of our analysis. We require that 
the ranges of
resonance electrocouplings extracted from the $\pi^+\pi^-p$ electroproduction
channel take into account both uncertainties in the measured differential cross sections
and in the JM model parameters. In order to provide a realistic
evaluation of resonance parameters, we abandoned
the traditional least-squares fit, since the parameters extracted in such a fit
correspond to a single presumed global minimum, while the experimental
data description achieved with others local minima may be equally good 
within the data error bars. Furthermore, the traditional evaluation of
fit-parameter uncertainties, based on the error propagation matrix, 
cannot be used for the same reason.

 A special 
 procedure was developed to obtain not only the best fit, 
 but also to establish bands 
 of computed cross sections that are compatible with the data within their 
 uncertainties. In the fit we vary simultaneously 
 non-resonant and resonant parameters of the JM model
 given in Tables~\ref{nstlist} and ~\ref{bckpar}, respectively,
 around their start values, employing unrestricted normal distributions. 
 The choice of the start
 parameters and the normal distribution $\sigma$-values employed for
 parameter variation will be further described below. For each trial set of the JM-model resonant and non-resonant
 parameters we  compute nine measured one-fold differential $\pi^+\pi^-p$ 
 cross sections, and 
 the $\chi^2$ per data point values ($\chi^2$/$d.p.$). The $\chi^2$/$d.p.$
 values were estimated in point-by-point comparisons between the  
 measured and computed 
 one-fold
 differential cross sections in all bins of $W$ and $Q^2$ 
 covered by the CLAS $\pi^+\pi^-p$ data of Ref. \cite{Fe09}. In the fit we 
 selected computed one-fold differential 
 cross
 sections closest to the data with  $\chi^2/d.p.$ less than a predetermined maximum 
 value $\chi^2_{max.}/d.p.$. 
 The values of  $\chi^2_{max.}/d.p.$ were obtained by requiring that the 
 computed cross
 sections with smaller $\chi^2/d.p.$ be within the data 
 uncertainties for the majority of the data points, based on point-by-point
 comparisons between the measured and the computed cross sections (see examples 
 in Figs.~\ref{9secnstbck},~\ref{9secnstbck1}). In this fit procedure we obtain the 
 $\chi^2/d.p.$ intervals within which the computed cross sections 
 describe the data equally well within the data uncertainties.

 The resonance parameters obtained from all these equally good fits are averaged 
 and their mean
 values are taken as the resonance parameters extracted from the data. 
 Dispersions in these parameters are taken as the uncertainties. 
 Our fitting 
 procedure  allows us to obtain more realistic uncertainties 
 of the fit parameters than from the usual least-squares method. 
 In this
 way we take into
 account both statistical uncertainties in the data and  
 systematic uncertainties imposed by the use of the JM reaction model.



We vary the parameters of the JM model, listed in
Table~\ref{bckpar} for the non-resonant 
mechanisms, which are relevant in describing the 
CLAS $\pi^+\pi^-p$ data \cite{Fe09}. They are limited to non-resonant $\pi \Delta$
sub-channels and direct $2\pi$-electroproduction.

\begin{table}
\begin{center}
\begin{tabular}{|c|c|}
\hline
                      & Ranges covered  \\
                      &  in variations of the  \\ 
Variable \, parameters &  start parameters,  \\
		       & \% from their values \\   		       		                   
\hline
Magnitude of the additional &     \\
contact term amplitude in the & 45.0                 \\
$\pi^- \Delta^{++}$ sub-channel &                \\  
\hline
Magnitude of the additional &     \\
contact term amplitude in the &  60.0        \\
$\pi^+ \Delta^{0}$ sub-channel &                \\
\hline
Magnitudes of six &    \\
2$\pi$ direct production&      30.0    \\
amplitudes                        &    \\
\hline
\end{tabular}
\caption{\label{bckpar} Variable parameters of non-resonant mechanisms
incorporated into the JM model \cite{Mo09}. The ranges in the table correspond
to the 3$\sigma$ areas around the start values of parameters.}
\end{center}
\end{table}

\begin{table}
\begin{center}
\begin{tabular}{|c|c|c|}
\hline
$N^{*}$ states   & Mass, & Total\; decay\; width,\; $\Gamma_{tot}$, \\
            &  (MeV)  &   (MeV)   \\
\hline
$P_{11}(1440)$ & 1430-1480 & 200-450   \\
$D_{13}(1520)$ & 1515-1530 & 100-150  \\
$S_{11}(1535)$ & 1510-1560 & 100-200  \\
\hline
\end{tabular}
\caption{\label{hadrrange} Allowed $N^*$ hadronic parameter variations 
in the fit of the CLAS $\pi^+\pi^-p$ electroproduction data \cite{Fe09}.}
\end{center}
\end{table}

The magnitudes of the additional contact terms in the $\pi^{-} \Delta^{++}$ and
$\pi^{+} \Delta^{0}$ isobar channels (row (b) in Fig.~\ref{mech_low_06}), 
as well as  the magnitudes of all direct
2$\pi$-production amplitudes (rows (c),(d) in Fig.~\ref{mech_low_06}), 
 are chosen as variable parameters. 
All these parameters
were determined in the fit to the CLAS data on $\pi^+\pi^-p$ electroproduction
in our previous analysis \cite{Mo09} without variation of the resonance parameters. In the CLAS data fit presented in this
paper, the values of the aforementioned non-resonant
parameters of the JM model are re-evaluated under simultaneous variation of:
\begin{itemize}
\item the magnitudes of 
additional contact-term amplitudes in the $\pi^{-} \Delta^{++}$ and
$\pi^{+} \Delta^{0}$ isobar channels (2 parameters per $Q^2$ bin);
\item the magnitudes of all direct
2$\pi$-production amplitudes (6 parameters per $Q^2$ bin);
\item the resonant
parameters listed in Table~\ref{nstlist}. The CLAS $\pi^+\pi^-p$ data \cite{Fe09} are mostly sensitive
to the electrocouplings of the $P_{11}(1440)$ and $D_{13}(1520)$ states 
(5 resonance electrocouplings per $Q^2$ bin), to the 
$\pi \Delta$ and $\rho p$ hadronic decay widths of these two resonances, and
to the $S_{11}(1535)$ state 
(6 parameters that remain the 
same in the entire $Q^2$ area covered by the measurements). 
\end{itemize}

Therefore, we consistently account for the
correlations between variations of the non-resonant and the resonant contributions 
while extracting 
the resonant parameters.

 The $W$-dependences of the magnitudes of the additional contact-term 
 amplitudes 
in the $\pi \Delta$
sub-channels and of the magnitudes of the direct  $2\pi$-production amplitudes 
are adopted from our previous analysis \cite{Mo09}. 
We apply
multiplicative factors to the magnitudes of the extra contact-term amplitudes 
and the 2$\pi$ direct-production amplitudes. 
The multiplicative factors are $W$-independent within any $Q^2$-bin, 
but they are fit to the data in each $Q^2$-bin independently.
 In this way we retain the smooth $W$-dependences of the non-resonant contributions
established in our previous analysis \cite{Mo09}.

We use two parameters for the variation of the magnitudes of additional 
contact-term amplitudes 
in the $\pi^-\Delta^{++}$ and $\pi^+\Delta^0$
isobar channels.  
However, the non-resonant parameters in the $\pi^+\Delta^0$ 
isobar channel have a rather small impact on the fit, since 
the non-resonant contributions of the $\pi^+\Delta^0$ isobar channel are approximately a 
factor 9 smaller than that of the $\pi^-\Delta^{++}$.

The parameters for the non-resonant Born terms in the $\pi \Delta$ sub-channels ((b)-row
in Fig.~\ref{mech_low_06}) include the $p \pi \Delta$ coupling and
cut-off for this hadron transition form factor, as well as the electromagnetic 
pion and nucleon form factors \cite{Ri00}. All these parameters 
are taken from previous studies 
of meson
photo-, electro-, and hadroproduction referred to in Ref. \cite{Mo09} and are kept fixed
for
the current fit.

Different assignments of final hadrons in $2\pi$-direct production
mechanisms, as shown in 
Fig.~\ref{mech_low_06} (rows (c),(d)), result in 12 different sub-processes. Our
previous analysis \cite{Mo09} demonstrated that the magnitudes of all these sub-processes are determined by six independent parameters.
In the current fit we vary these six parameters of the direct 2$\pi$-production amplitudes.

In the fitting procedure described above, nine one-fold $\pi^+\pi^-p$ differential
 cross sections are computed with non-resonant variable 
parameters of the JM model obtained by employing unrestricted normal distributions 
around their start values 
with $\sigma$-parameters of 10\%-20\% of their start values 
(see Table ~\ref{bckpar}). In this way we explore mostly the range 
of $\approx$ 3$\sigma$ around the start parameter values. These ranges for
non-resonant 
parameter variations in the JM model are  shown in Table~\ref{bckpar} as a percentage of their start values. 

In this fit we also vary the $\gamma_{v}pN^*$
electrocouplings and the $\pi \Delta$ and $\rho p$ hadronic partial decay widths 
of the 
$P_{11}(1440)$ and $D_{13}(1520)$ resonances 
around their start
values. 
The start 
values of the $P_{11}(1440)$ and
$D_{13}(1520)$ electrocouplings are determined by interpolating 
the results from the analysis \cite{Az051} of the CLAS data on $N\pi$ 
electroproduction off protons into the range 0.4 GeV$^2$ $<$ $Q^2$ $<$ 0.6 GeV$^2$ and extrapolating these
results into the $Q^2$-area from 0.25 GeV$^2$ to 0.4 GeV$^2$. 
The electrocouplings 
of the $P_{11}(1440)$ and
$D_{13}(1520)$ resonances
 are varied employing normal distributions with
$\sigma$ parameters equal to 30\% of their start values. There are no
restrictions on the minimum or maximum trial electrocoupling values. 
The normal distributions allow us to explore mostly 
the area of $\approx$ 3$\sigma$ around their start values or 
90\% around electrocoupling start values.

The $\pi^+\pi^-p$ electroproduction channel has also some 
sensitivity to the $S_{11}(1535)$ state, 
which couples dominantly  to the $N\pi$ and $N\eta$ final states. The 
$S_{11}(1535)$ electrocouplings were taken from 
the CLAS analysis
of $N \pi$ electroproduction \cite{Az09} and varied strictly inside 
the uncertainties reported in that
paper.

The $\pi \Delta$ and $\rho p$ hadronic decay widths of the $P_{11}(1440)$,
$D_{13}(1520)$, and $S_{11}(1535)$ resonances are varied around their start values taken from previous 
analyses of the CLAS double pion electroproduction data \cite{Az05,Mo06,Mo06-1} in ranges restricted 
by the total
$N^*$ decay widths and their uncertainties, as shown in Table~\ref{hadrrange}. 
The total $N^{*}$ decay widths were 
obtained by summing 
the partial widths over all decay channels. Partial hadronic 
decay widths to all final states other than $\pi \Delta$ and $\rho p$ are
computed as the products of RPP \cite{PDG10} values of $N^*$ total decay widths and 
branching fractions for decays to particular hadronic final states, that are taken from analyses \cite{Man92} of hadroproduction experiments.  
We varied the $\pi \Delta$ and $\rho p$ hadronic decay widths of the 
$P_{11}(1440)$,
$D_{13}(1520)$, and $S_{11}(1535)$ resonances simultaneously with their 
masses, keeping the hadronic 
$N^*$ parameters independent of $Q^2$. Accounting for the correlations of the $N^*$ 
electromagnetic and hadronic decay
parameters in a combined variation is important for a credible extraction of the resonance parameters and, in particular, for the evaluation 
of their uncertainties.

The here developed fit procedure allows us to determine the $\pi \Delta$ and $\rho p$ hadronic decay widths 
of the $P_{11}(1440)$, $D_{13}(1520)$, and $S_{11}(1535)$. The resonance total decay widths are affected 
by
the variation of their $\pi \Delta$ and $\rho p$ partial decay widths. Instead, when varying resonance electrocuplings, the $N^*$ 
total decay widths of resonances are kept almost unchanged. Resonance hadronic decay
amplitudes are independent of $Q^2$, while resonance electrocouplings represent the functional dependencies on the photon virtualities. 
These distinctive features allow us to disentangle resonance electromagnetic
and hadronic decay amplitudes in the fit of the CLAS data \cite{Fe09} and to obtain  
the $\pi \Delta$ and $\rho p$ hadronic decay
widths of the $P_{11}(1440)$, $D_{13}(1520)$, and $S_{11}(1535)$.

To account for the contributions from the tails of higher mass
resonances, electrocouplings of all $N^*$ states marked in Table~\ref{nstlist} 
as ``var" are varied around their start values. For resonances with masses 
above 1.6 GeV, except the $S_{11}(1650)$ state, the start values 
for electrocouplings are taken from the results of 
previous analyses  \cite{Az05,Mo06} of $\pi^+\pi^-p$
electroproduction data \cite{Ri03} after the extrapolation 
into the $Q^2$ range of the current analysis (0.25-0.6 GeV$^2$). 
The $S_{11}(1650)$ state may have a more pronounced impact on the extraction of
the $P_{11}(1440)$ and $D_{13}(1520)$ resonance parameters, since
 the $S_{11}(1535)$ and $S_{11}(1650)$ states are mixed in the 
 dressed resonance propagator of the unitarized BW ansatz employed in the JM model.
For this reason we prefer to have more accurate start
values for the $S_{11}(1650)$ electrocouplings. They are computed within the 
framework of the single quark transition model (SQTM) \cite{Bu03} from electrocouplings of 
the $S_{11}(1535)$ state that are available from analyses of $N\pi$
and $N\eta$ electroproduction data \cite{Bu11b}.
Electrocouplings of the $D_{13}(1700)$ and 
$D_{33}(1700)$ resonances are kept
fixed at their start values, since the masses of these resonances are far 
outside of
the $W$ area covered by the CLAS $\pi^+\pi^-p$ electroproduction data \cite{Fe09} $W$ $<$ 1.6 GeV.

The $\pi \Delta$ and $\rho p$ partial hadronic decay widths for $N^{*}$ states
other than the $P_{11}(1440)$,
$D_{13}(1520)$, and $S_{11}(1535)$ resonances are taken from previous 
analyses of the CLAS double pion electroproduction data \cite{Az05,Mo06,Mo06-1}.  They are in  a reasonable agreement
with the values reported in the RPP \cite{PDG10}. The respective branching fractions 
are listed in
Table~\ref{nstlist}.

\begin{table}
\begin{center}
\begin{tabular}{|c|c|c|c|}
\hline
 $Q^{2}$ bins, & $W$ intervals, & Number of fitted  &
$\chi^2/d.p.$  intervals\\
 (GeV$^2$)&  (GeV)  & points   & for selected 
  \\
 & & & cross sections \\ 
\hline
0.25-0.30 & 1.44-1.56 & 378 & $~~~$  \\
0.30-0.35 & 1.41-1.54 & 378 & 2.66-2.74  \\
0.35-0.40 & 1.41-1.54 & 378 & $~~~$  \\
\hline
0.40-0.45 & 1.39-1.51 & 373 & 1.87-2.04  \\
0.45-0.50 & 1.39-1.49 & 309 & $~~~$  \\
\hline
0.50-0.55 & 1.39-1.51 & 360 & 1.57-1.75  \\
0.55-0.60 & 1.34-1.44 & 279 & $~~~$  \\
\hline
\end{tabular}
\caption{\label{fitqual} Quality of the CLAS $\pi^+\pi^-p$  
fit within the framework 
of the JM model \cite{Mo09}. Horizontal lines separate the three 
$Q^2$-intervals, in which the fits were carried-out independently.}
\end{center}
\end{table}

We
 fit the CLAS data \cite{Fe09} consisting of nine differential cross sections 
 of the
 $ep \rightarrow e'p'\pi^+\pi^-$ electroproduction reaction 
 in all bins of $W$ and $Q^2$ in 
 the kinematic region: 1.35 GeV $<$ $W$ $<$ 1.57 GeV and 0.25 GeV$^2$ $<$ $Q^2$
 $<$ 0.6 GeV$^2$ within the framework of the fit procedure described above. Three
 intervals of $Q^2$ separated in Table~\ref{fitqual} by horizontal solid lines are fit independently.   
 The $\chi^2/d.p.$ intervals, that correspond to 
 equally good data description
 within the error bars are shown in Table~\ref{fitqual}. Their values
 demonstrate the quality of the CLAS $\pi^+\pi^-p$ data 
 description achieved in the fits. Examining the description 
 of the nine one-fold
 differential cross sections, we found that the $\chi^2/d.p.$ values 
 are determined mostly by the random deviations 
 of some experimental data points from the bunches of computed fit 
 cross sections. There are no discrepancies in describing the
 shapes of the differential cross sections, which would manifest themselves 
 systematically in neighboring 
 bins of $W$ and $Q^2$. 
 Typical examples  for $W$=1.51 GeV and neighboring  $Q^2$ intervals
 centered at 0.38 GeV$^2$ and 0.42 GeV$^2$  are 
 shown in Fig.~\ref{9secnstbck} and Fig.~\ref{9secnstbck1}, respectively.

The sets of computed differential cross sections with  $\chi^2/d.p.$ values within 
the intervals given in the Table~\ref{fitqual} offer the best data descriptions achievable 
within the framework of 
the JM model.  
The minimal values of $\chi^2/d.p.$  in each interval represent the global minima 
of these fits. 
We found that the increases of the $\chi^2/d.p.$ values within the intervals listed in the Table~\ref{fitqual} change 
the computed 
differential cross sections, but still keep them inside the data uncertainties, offering equally good data descriptions in 
all these fits with different sets of JM model parameters. 
The mean values of the resonance parameters from these sets and their dispersions (statistical r.m.s.) are determining the resonance parameters and 
their uncertainties, as it was described at the beginning of this Section.

Since only statistical data uncertainties are used 
 in the computation of the $\chi^2/d.p.$ values listed in the Table~\ref{fitqual}, 
 we concluded 
 that a reasonable data description was
 achieved.  
 The $\chi^2/d.p.$ values of our fits 
 are comparable with those obtained in the fit 
 of the CLAS $N\pi$ electroproduction data published in \cite{Az09}, as well as with those obtained 
 in the MAID analysis \cite{Tia07}.


\section{Evaluation of $\gamma_{v}pN^*$ electrocouplings and resonance hadronic decay parameters}
\label{nstarelectrocoupl}

The extraction of resonance electrocouplings using the JM model 
relies on fitting resonant 
and non-resonant contributions to measured differential cross sections. Therefore, we first have to check the quality of the 
separation between  resonant and
non-resonant contributions achieved in the data fit.


\subsection{Separation of resonant and non-resonant contributions to the $\pi^+\pi^-p$ cross sections}
\label{nstarbackgr} 
\begin{figure*}[btp]
\includegraphics[width=18cm,height=20cm]
{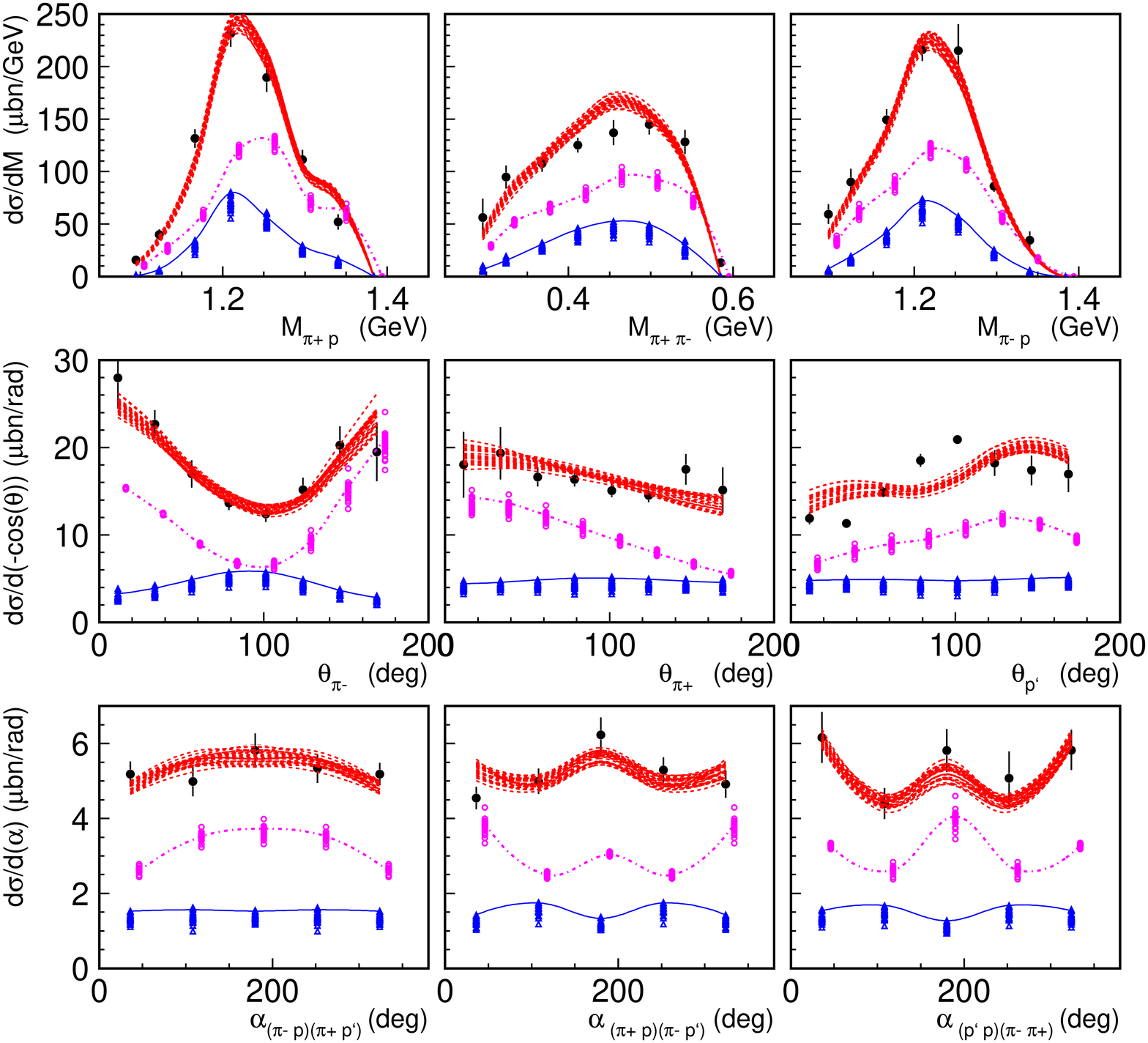}
\caption{\small(color online) Resonant (blue triangles) and non-resonant (green open circles) contributions 
to the differential cross sections (red lines) obtained from the CLAS data \cite{Fe09}, fit within the framework of 
the JM model at
$W$= 1.51 GeV, $Q^2$=0.38 GeV$^2$. The solid blue and dotted-dashed green lines stand for the resonant and non-resonant contributions, respectively, which were computed  for minimal $\chi^2/d.p.$ achieved in the data fit.  The points for resonant and non-resonant contributions are shifted on each panel for better visibility. Dashed lines show selected  
fits.}
\label{9secnstbck}
\end{figure*}

\begin{figure*}[btp]
\includegraphics[width=18cm,height=20cm]
{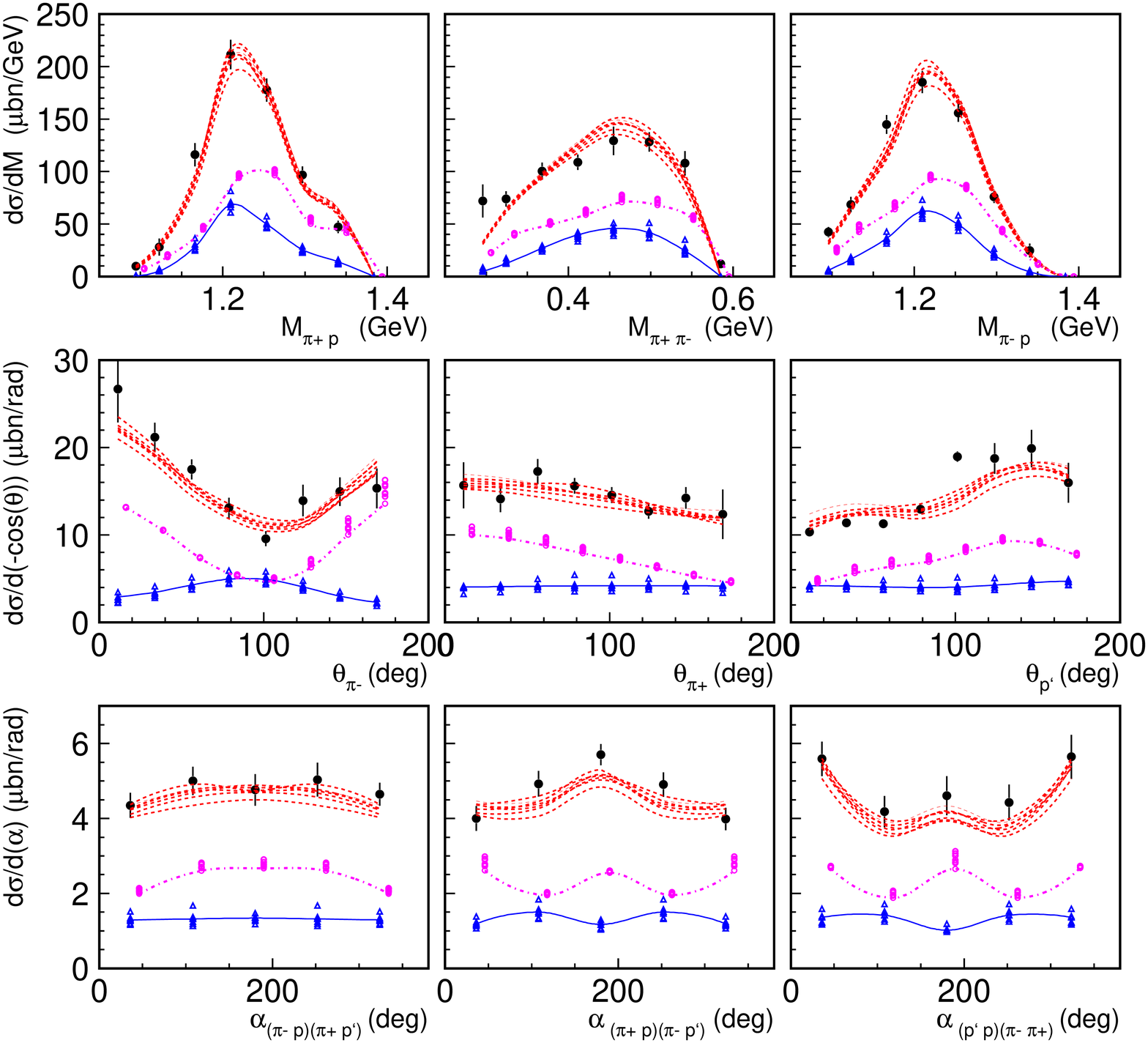}
\caption{\small(color online) The same as in Fig.~\ref{9secnstbck} at
$W$= 1.51 GeV, $Q^2$=0.43 GeV$^2$.}
\label{9secnstbck1}
\end{figure*}

With the parameters determined from the fits, we now use the JM model 
to evaluate 
the contributions from resonant and non-resonant parts to the cross sections. 
This is done by computing the nine differential 
cross sections
without the resonant parts  and with only the resonant parts for all 
 trial differential cross sections of the JM model selected in the fit. 
In this way we also determine the ranges of the 
resonant and non-resonant contributions to the cross sections 
as imposed by the uncertainties of the experimental
data. These ranges account for the uncertainties of the non-resonant
parameters listed in Table ~\ref{bckpar} and for all correlations between
resonance and non-resonant amplitudes. Therefore, we obtain reliable 
estimates for the uncertainties of the resonant and non-resonant contributions to the differential cross sections. 
Examples of the separated resonant and non-resonant contributions 
for two particular $W$ and $Q^2$ bins are shown 
in Figs.~\ref{9secnstbck} and~\ref{9secnstbck1}. The interference 
between the resonant and non-resonant amplitudes is clearly seen in all
angular distributions. The differences between the fitted
$\pi^+\pi^-p$ differential cross sections and their non-resonant parts are 
larger than the resonant contributions for a majority of the data points in 
the angular distributions of the final hadrons, offering clear evidence for  
the interference between resonant and non-resonant
amplitudes. This interference amplifies
the $N^*$ contributions to all CM-angular distributions 
d$\sigma$/d$(-cos(\theta_{i}))$ ($i=\pi^+, \pi^-, p$)
of the final hadrons, which improves the sensitivity of the fit to the 
resonance parameters.  Furthermore, the shapes 
of the resonant and non-resonant  differential cross sections  
are rather different, especially for the final-state  
angular distributions. Substantial differences in the shapes of the 
resonant/non-resonant contributions and their interference allow us 
to isolate the resonant contribution in a
combined fit of all nine one-fold differential cross sections despite the relatively 
small resonant contributions to the fully integrated cross sections. 
In our previous analysis \cite{Fe09} we found resonant contributions to the fully 
integrated cross sections from 10\% to 30\%.

The examples of Figs.~\ref{9secnstbck} and~\ref{9secnstbck1} demonstrate 
that the uncertainties  
of the resonance parts 
are comparable with those of the experimental
data. This is further evidence for the credible separation 
between the resonant and non-resonant contributions.
Any ambiguities in the evaluation of these two contributions would
result in larger uncertainties for the resonant and/or non-resonant
parts derived from the fit than the uncertainties of the original data. But this is not the case in the entire kinematical area covered by the
CLAS $\pi^+\pi^-p$ electroproduction measurements of Ref.~\cite{Fe09}. 
Therefore, these data provide enough constraints in order to determine the resonant contributions to all differential cross sections.



\subsection{$N^*$ parameters from the fit of the $\rm \pi^+\pi^-p$ electroproduction cross sections}
\label{resonances}

 The procedure described in Section~\ref{fit} allows us to extract the $\gamma_{v}pN^*$ electrocouplings and their uncertainties. 
A special approach was developed for the evaluation 
of the $P_{11}(1440)$ electrocouplings. According to the analysis \cite{Az09} of
the CLAS $N\pi$ electroproduction data, the $A_{1/2}$ electrocoupling of this
resonance changes sign between $Q^2$ of 0.40 GeV$^2$ and 0.45 GeV$^2$. 
Here the $A_{1/2}$ electrocoupling is close to zero and is an order of 
magnitude smaller 
than $S_{1/2}$. The $A_{1/2}$ variations computed as a 
percentage of the start value, which is close to
zero, become too small. For realistic uncertainty 
estimates we varied $A_{1/2}$ for 0.4 GeV$^2$ $<$ $Q^2$ $<$ 0.5
GeV$^2$ in the ranges shown in  Table~\ref{range}. 
By varying $A_{1/2}$ inside these widened ranges, we scanned the trial
values comparable with those of $S_{1/2}$, as they were obtained in the
analysis \cite{Az09} of the CLAS $N\pi$ electroproduction data.  
We refit the CLAS data of Ref.~\cite{Fe09} on $\pi^+\pi^-p$ electroproduction by varying
$A_{1/2}$, as described above, keeping the variation of all other resonant and non-resonant parameters as described in Section~\ref{fit}.

\begin{table}
\begin{center}
\begin{tabular}{|c|c|}
\hline
$Q^2$ bins,         & Ranges covered  \\
(GeV$^2$)                    &     in variations of $A_{1/2}$                                       \\ 
                    &  electrocoupling of the \\
		      &    $P_{11}(1440)$ resonance, \\ 
		     & (10$^{-3}$ GeV$^{-1/2}$) \\   		       		                   
\hline
0.40-0.45            & -20 \, \,-\,\, +5     \\

0.45-0.50            &  -5 \, \,-\,\, 20   \\
\hline
\end{tabular}
\caption{\label{range} Variation range for the 
$P_{11}(1440)$ $A_{1/2}$ electrocoupling at photon
virtualities where this electrocoupling changes sign.}
\end{center}
\end{table}

\begin{figure}[htp]
\begin{center}
\includegraphics[width=8.cm]{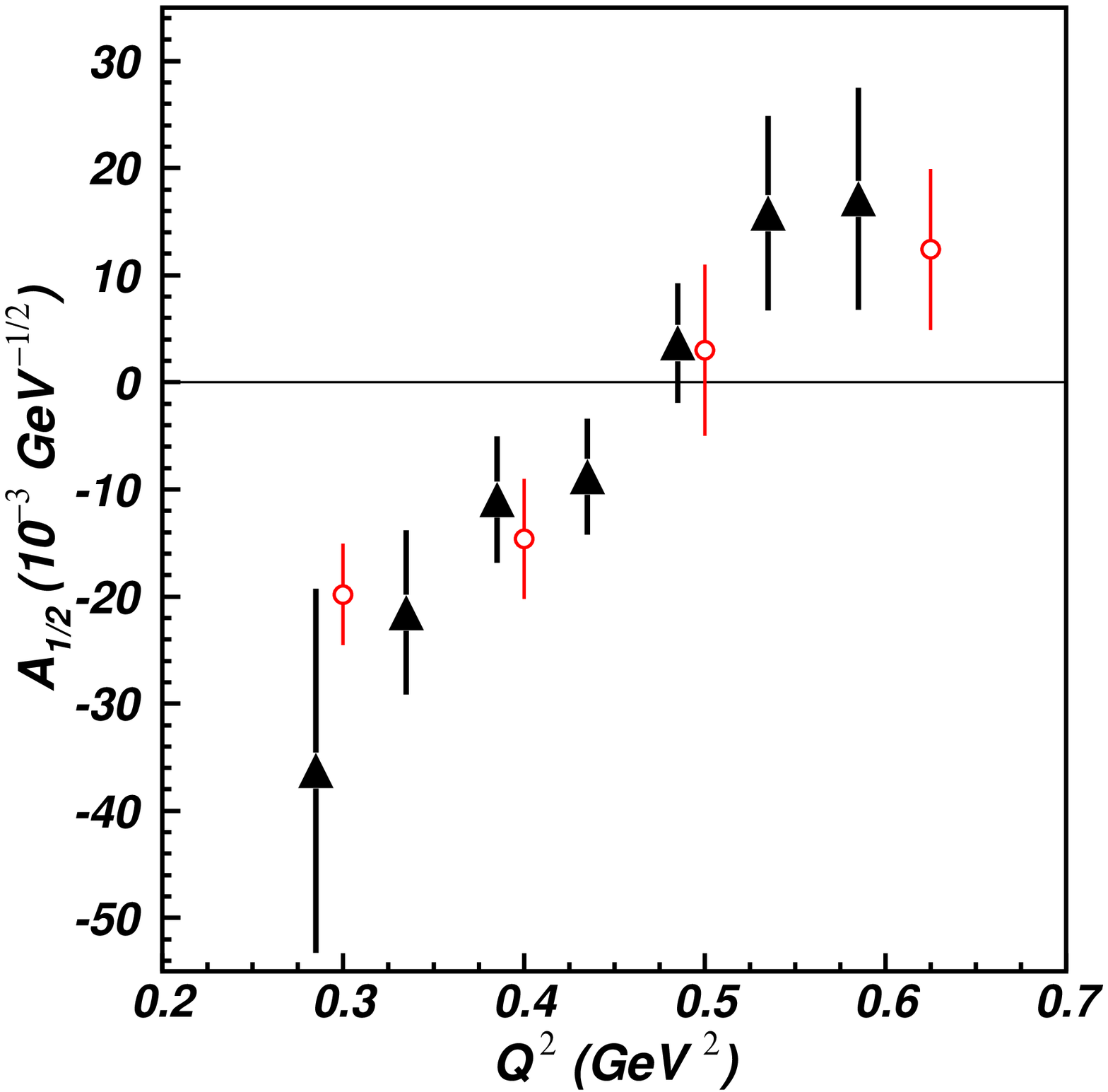}
\includegraphics[width=8.cm]{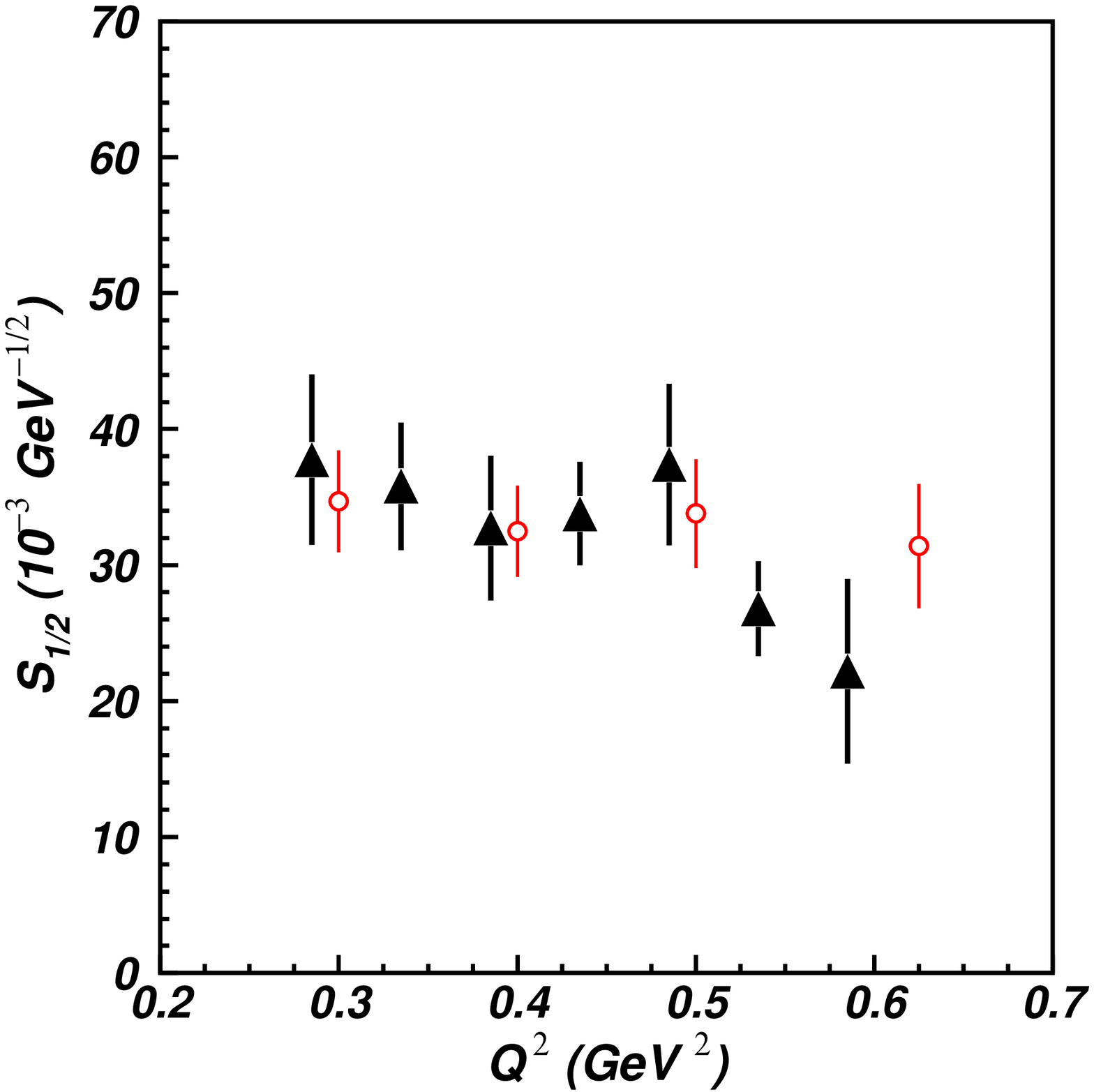}

\caption{\small(color online) Electrocouplings and full error bars of the $P_{11}(1440)$ state. 
The results from the analysis of 
the CLAS $\pi^+\pi^-p$ electroproduction data \cite{Fe09} are shown by triangles. 
Electrocouplings from the analysis of the $N\pi$ electroproduction 
data \cite{Az09} are shown by circles.}  
\label{p11_23}
\end{center}
\end{figure}

In order to compare our results on the $P_{11}(1440)$ and $D_{13}(1520)$ 
electrocouplings in the $\pi^+\pi^-p$ 
electroproduction channel with their values from the analysis of 
$N\pi$ electroproduction \cite{Az09}, we have to use in both of the exclusive
electroproduction channels common branching
fractions for decays of these resonances to the $N\pi$ and $N\pi\pi$ 
final states. According to the RPP \cite{PDG10}, the sum of the branching 
fractions into the $N\pi$ and $N\pi\pi$ final states accounts for almost 100\% of the total decay widths of the  $P_{11}(1440)$ and $D_{13}(1520)$ states.
In our analysis the branching fractions for $P_{11}(1440)$ and $D_{13}(1520)$ 
resonance decays 
to the $N\pi$ final state $BF(N\pi)$ 
were taken from the previous CLAS studies of $N\pi$ electroproduction data 
\cite{Az09}, since the $N\pi$ exclusive electroproduction channels are 
most sensitive to contributions from the 
$P_{11}(1440)$ and $D_{13}(1520)$ resonances. The unitarity condition
allows us to estimate the branching fraction $BF(N\pi\pi)_{corr}$ value as:
\begin{equation}
\label{bnpipi}
BF(N\pi\pi)_{corr}=1-BF(N\pi).
\end{equation} 
For these resonance decays to the $N\pi\pi$ final states it turns out that 
the estimated branching fractions  $BF(N\pi\pi)_{corr}$ 
from Eq.~(\ref{bnpipi}) 
  are slightly ($<$ 10\%) different with respect to those 
 obtained from the $\pi^+\pi^-p$ fit ($BF(N\pi\pi)_{0}$). Therefore, 
we multiplied
the $\pi \Delta$ and $\rho p$ hadronic decay widths of
 $P_{11}(1440)$ and $D_{13}(1520)$ from the $\pi^+\pi^-p$ fit by the ratio
$\frac{BF(N\pi\pi)_{corr}}{BF(N\pi\pi)_{0}}$ in order to make them consistent
with the unitarity condition of Eq.~(\ref{bnpipi}).

\begin{figure}[htp]
\begin{center}
\includegraphics[width=7.5cm]{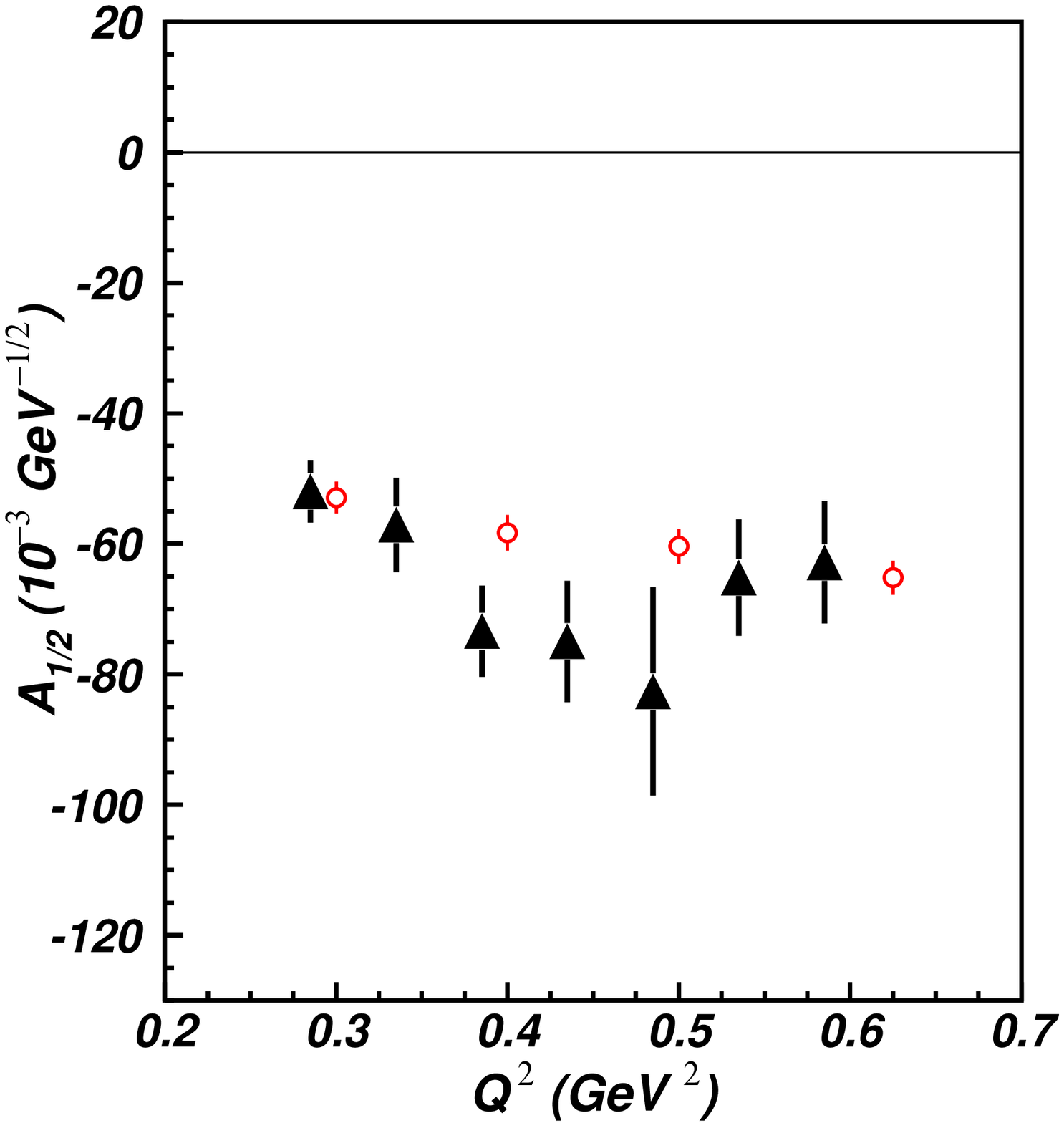}
\includegraphics[width=7.5cm]{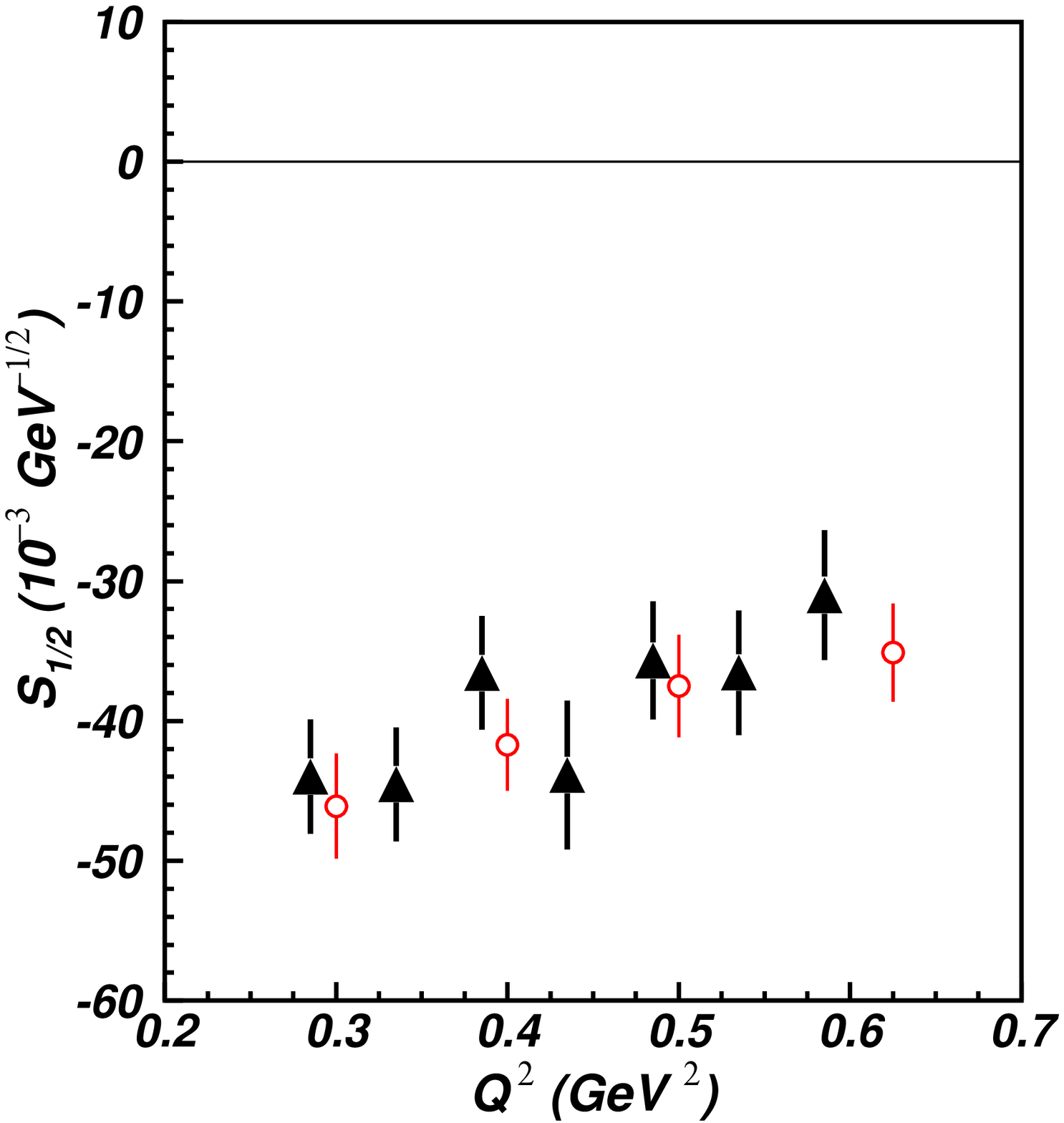}
\includegraphics[width=7.5cm]{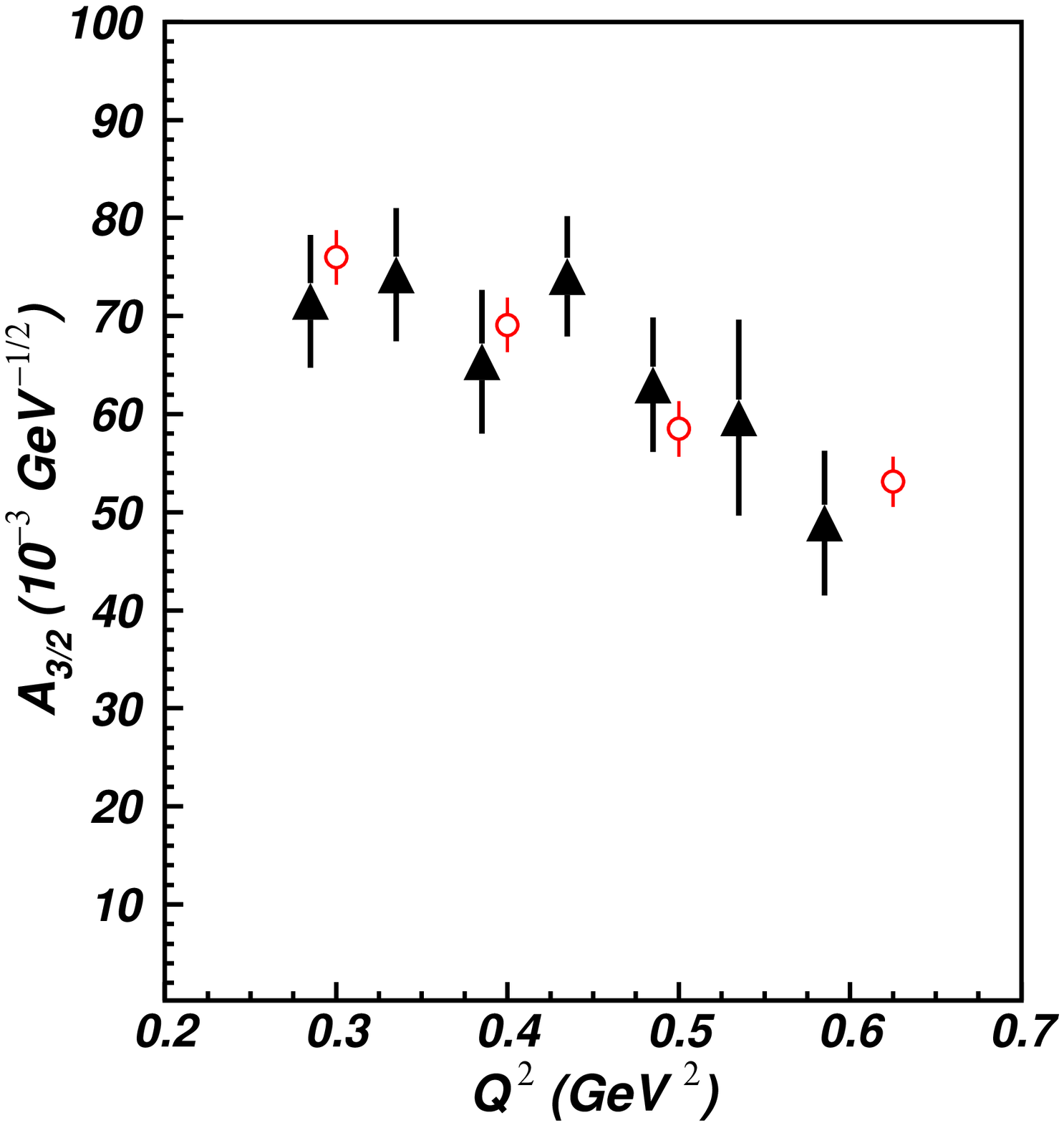}
\caption{\small(color online) Electrocouplings and full error bars of the $D_{13}(1520)$ state. 
This analysis of 
the CLAS $\pi^+\pi^-p$ electroproduction data \cite{Fe09} is shown by triangles. 
Electrocouplings from the analysis of the $N\pi$ electroproduction 
data \cite{Az09} are shown by circles.}  
\label{d13_23}
\end{center}
\end{figure}


\begin{figure}[htp]
\begin{center}
\includegraphics[width=8cm]{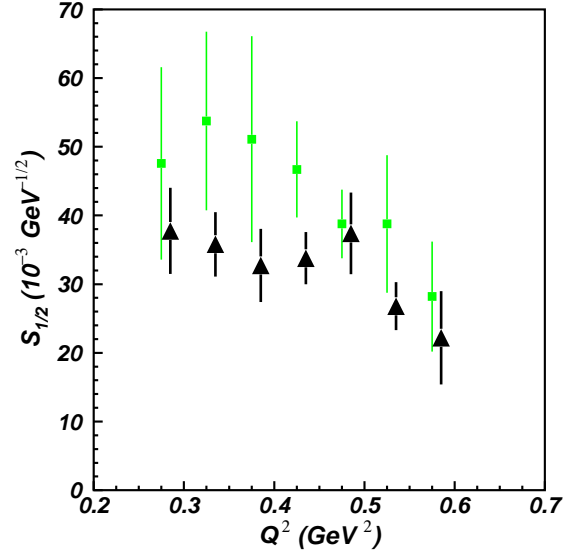}

\caption{\small(color online) Comparison between the final results on $S_{1/2}$ electrocouplings 
for the $P_{11}(1440)$ from this analysis of CLAS $\pi^+\pi^-p$ electroproduction data \cite{Fe09} (triangles) 
with the values 
obtained by employing a regular BW ansatz for the resonant amplitudes and with the contributions from the $S_{11}(1535)$ 
state turned off (squares).}  
\label{p11_s12}
\end{center}
\end{figure}

Consequently, the $P_{11}(1440)$ and $D_{13}(1520)$ electrocouplings obtained in 
our analysis
are multiplied by correction factors 
\begin{equation}
\label{bnpipi1}
C_{hd}=\sqrt{\frac{BF(N\pi\pi)_{0}}{BF(N\pi\pi)_{corr}}},
\end{equation} 
in order to keep the resonant parts and the full computed differential $\pi^+\pi^-p$ 
cross sections unchanged under the re-scaling 
of resonance hadronic decay
parameters described above.

\begin{table}
\begin{center}
\begin{tabular}{|c|c|c|}
\hline
Q$^{2}$, & $A_{1/2}$,          &  $S_{1/2}$,           \\
(GeV$^2)$            & (10$^{-3}$ GeV$^{-1/2})$  &  (10$^{-3}$ GeV$^{-1/2})$  \\
\hline
$0.28$ & $-36.3 \pm 17.0$ & $37.8 \pm 6.3$  \\
$0.33$ & $-21.5 \pm 7.7$ & $35.8 \pm 4.7$ \\
$0.38$ & $-10.9 \pm 5.9$ & $32.7 \pm 5.3$ \\
$0.43$ & $-8.8 \pm 5.4$ & $33.8 \pm 3.8$ \\
$0.48$ & $3.7 \pm 5.9$ & $37.4 \pm 5.9$ \\
$0.53$ & $15.8 \pm 9.1 $ & $26.8 \pm 3.5$  \\
$0.58$ & $17.2 \pm 10.4$ & $22.2 \pm 6.8$ \\
\hline
\end{tabular}
\caption{\label{ecp11} Electrocouplings of the $P_{11}(1440)$ resonance 
determined from this analysis of
$\pi^+\pi^-p$ electroproduction off protons within the framework of the 
JM model \cite{Mo09}.}
\end{center}
\end{table}

\begin{table}
\begin{center}
\begin{tabular}{|c|c|c|c|}
\hline
Q$^{2}$, & $A_{1/2}$,          & $S_{1/2}$,         & $A_{3/2}$, \\
 (GeV$^2)$            & (10$^{-3}$ GeV$^{-1/2})$  & (10$^{-3}$ GeV$^{-1/2}$ & (10$^{-3})$ GeV$^{-1/2})$ \\
\hline
$0.28$ & $-51.9 \pm 4.8$ & $-44.0 \pm 4.1$ & $71.5 \pm 6.8$ \\
$0.33$ & $-57.1 \pm 7.2$ & $-44.5 \pm 4.1$ & $74.2 \pm 6.8$ \\
$0.38$ & $-73.4 \pm 7.0$ & $-36.5 \pm 4.1$ & $65.3 \pm 7.3$ \\
$0.43$ & $-75.0 \pm 9.3$ & $-43.9 \pm 5.3$ & $74.1 \pm 6.1$ \\
$0.48$ & $-82.6 \pm 15.9$ & $-35.7 \pm 4.2$ & $63.0 \pm 6.8$ \\
$0.53$ & $-65.2 \pm 8.9 $ & $-36.6 \pm 4.5$ & $59.6 \pm 10.0$ \\
$0.58$ & $-62.2 \pm 9.3$ & $-31.0 \pm 4.6$ & $48.9 \pm 7.4$ \\
\hline
\end{tabular}
\caption{\label{ecd13} Electrocouplings of the $D_{13}(1520)$ resonance determined
 from this analysis of
$\pi^+\pi^-p$ electroproduction off protons within the framework of the JM model \cite{Mo09}.}
\end{center}
\end{table}

\begin{table*}
\begin{center}
\begin{tabular}{|c|c|c|}
\hline
Parameter & Analysis of the CLAS  & RPP \\
   &  $\pi^+\pi^-p$ data &   \\
\hline
Breit-Wigner mass, MeV & 1458 $\pm$ 12 & 1420-1470 ($\approx$ 1440) \\
Breit-Wigner width, MeV & 363 $\pm$ 39 & 200-450 ($\approx$ 300) \\
$\pi \Delta$ partial decay width, MeV & 142 $\pm$ 48 & \\
$\pi \Delta$ BF,  & 23\%-58\% &  20\%-30\%     \\
$\rho ~p$ partial decay width, MeV & 6.2 $\pm$ 4.1 &  \\
$\rho ~p$ BF & $<$ $~$2.0\% &  $<$ $~$8.0\%     \\
\hline
\end{tabular}
\caption{\label{hpp11} Hadronic parameters of the $P_{11}(1440)$ resonance determined from the CLAS data \cite{Fe09} on
$\pi^+\pi^-p$ electroproduction off protons within the framework of the JM model \cite{Mo09} and from the RPP \cite{PDG10}.}
\end{center}
\end{table*}

\begin{table*}
\begin{center}
\begin{tabular}{|c|c|c|}
\hline
Parameter & Analysis of the CLAS  & RPP \\
          &  $\pi^+\pi^-p$ data   &   \\
\hline
Breit-Wigner mass, MeV & 1521 $\pm$ 4 & 1515-1525 ($\approx$ 1520) \\
Breit-Wigner width, MeV & 127 $\pm$ 4 & 100-125 ($\approx$ 115) \\
$\pi \Delta$ partial decay width, MeV & 35 $\pm$ 4 &  \\
$\pi \Delta$ BF & 24\%-32\% &  15\%-25\%     \\
$\rho ~p$ partial decay width, MeV & 16 $\pm$ 5 &  \\
$\rho ~p$ BF  & 8.4\%-17\% &   15\%-25\%     \\
\hline
\end{tabular}
\caption{\label{hpd13} Hadronic parameters of the $D_{13}(1520)$ resonance determined from the CLAS data \cite{Fe09} on
$\pi^+\pi^-p$ electroproduction off protons within the framework of the JM model \cite{Mo09} and from the RPP \cite{PDG10}.}
\end{center}
\end{table*}



The $P_{11}(1440)$ and $D_{13}(1520)$ electrocouplings derived from 
the fit of $\pi^+\pi^-p$ electroproduction data are presented in Tables~\ref{ecp11} 
and ~\ref{ecd13}. The masses and hadronic 
decay parameters are shown in Tables~\ref{hpp11} and ~\ref{hpd13}. 
The hadronic resonance parameters are taken as the average of their values 
obtained in independent fits in the three 
$Q^2$ intervals shown in Table~\ref{fitqual}.


In Figures~\ref{p11_23} and~\ref{d13_23} we compare the results of this analysis with 
results of the analysis of CLAS $N\pi$ electroproduction data \cite{Az09}. Furthermore, there are also 
the  $P_{11}(1440)$ and $D_{13}(1520)$ electrocouplings available from the MAID partial wave analysis of 
$N\pi$
electroproduction at $Q^2$ $<$ 0.6 GeV$^2$ \cite{Tia12,Tia09,Tia07}.  
Our analysis confirms the
sign change of the $A_{1/2}$ electrocoupling for the $P_{11}(1440)$, 
first observed in the $N\pi$ channels. We
found the zero-crossing to be between $Q^2$=0.4 GeV$^2$ and 0.45 GeV$^2$ in agreement with the $N\pi$ analysis \cite{Az09}. The electrocouplings for the $P_{11}(1440)$ and $D_{13}(1520)$ are also consistent within their uncertainties 
with the CLAS analysis of the $N\pi$ electroproduction  \cite{Az09} 
with the exception of the $A_{1/2}$ amplitude of $D_{13}(1520)$,  
where we see a difference in the range of 0.4 GeV$^2$ $<$ $Q^2$ $<$ 0.5 GeV$^2$, which is slightly larger 
than a standard deviation. In this range of $Q^2$ two different analyses of the 
$N\pi$ exclusive channels the CLAS \cite{Az09} and the MAID \cite{Tia12}, give also slightly different
values of the $A_{1/2}$ electrocoupling for the $D_{13}(1520)$. The MAID results  are close to our extraction at 0.4 GeV$^2$ $<$ $Q^2$ $<$ 0.5 GeV$^2$, but start to deviate at larger photon virtualities. Instead, for the other electrocouplings of $D_{13}(1520)$, both the CLAS \cite{Az09} and the MAID \cite{Tia12} analyses of the $N\pi$ channels are consistent with the
results of this analysis.

The unitarization of the resonant amplitudes and 
the contribution from the $S_{11}(1535)$ state 
have a negligible impact on $A_{1/2}$ for $P_{11}(1440)$ and on 
all $D_{13}(1520)$ electrocouplings. However,
both factors affect the $S_{1/2}$ electrocouplings of $P_{11}(1440)$. 
Figure~\ref{p11_s12} shows a comparison 
of the $P_{11}(1440)$ $S_{1/2}$ electrocouplings obtained with 
the unitarized BW ansatz incorporating the $S_{11}(1535)$ contributions
versus the regular 
BW ansatz without the $S_{11}(1535)$. The implementation of the  
$S_{11}(1535)$ contributions imposes additional constraints on the ranges of
the $P_{11}(1440)$ $S_{1/2}$ electrocouplings that correspond to the 
computed differential cross
sections, which are consistent with the data, allowing us to improve the 
uncertainties in the extraction of this electrocoupling.

Consistent results on the electrocouplings of the $P_{11}(1440)$ and 
$D_{13}(1520)$
resonances, which are available for the first time from independent analyses 
of the  major $N\pi$ and
$\pi^+\pi^-p$ exclusive channels provide evidence for the reliable extraction of these
fundamental quantities from the experimental data. Furthermore, this
agreement strongly supports the reaction models employed for the analyses of the 
$N\pi$ \cite{Az09} and $\pi^+\pi^-p$ \cite{Mo09} electroproduction data.



The Breit-Wigner masses and mean values for the total decay widths 
of the $P_{11}(1440)$ and $D_{13}(1520)$, derived 
from the $\pi^+\pi^-p$ data and listed in Tables~\ref{hpp11} and ~\ref{hpd13}, are in a good
agreement with those reported in the RPP \cite{PDG10}. Uncertainties in the total resonance decay widths  were obtained by
varying the $\pi \Delta$ and $\rho p$ partial decay widths, while the decay 
width to the $N\pi$  final state remained 
fixed at the
values taken from Ref.~\cite{Az09}. In comparison with the results in the RPP \cite{PDG10}, 
our analysis gives somewhat larger branching fractions to the $\pi \Delta$ final states for the hadronic decays
of both the $P_{11}(1440)$ and $D_{13}(1520)$ states, and a smaller hadronic decay to the
 $\rho p$ final state. 

Both $\gamma_{v}pN^*$ 
electrocouplings and resonance 
hadronic decay widths are
obtained at the resonant point on the real energy
axis $W=M_{r}$, using the relativistic Breit-Wigner parametrizations of the resonant 
amplitudes. All resonance parameters determined in our approach incorporate 
$all$ relevant contributions 
to the $N^*$ structure. The extraction from the data of the individual contributions of the  quark core and the meson-baryon dressing is outside of our scope. 
Our results can be 
compared directly with those determined from other meson
electroproduction channels, where the BW ansatz for resonant amplitudes is used. 
For comparisons
with results obtained from models that use other ways to 
describe the resonant contributions, 
the resonant cross sections at the resonant points can be compared.



\section{The impact on the studies of
 resonance structure from the CLAS results 
 on the $P_{11}(1440)$ and $D_{13}(1520)$ electrocouplings and hadronic 
parameters.}
In this section we discuss the impact of the CLAS data on 
the $P_{11}(1440)$ and $D_{13}(1520)$
electrocouplings and hadronic parameters determined from the independent analyses of $N\pi$ and $\pi^+\pi^-p$ electroproduction off protons
on the contemporary understanding of the structures these states at 
various distances.

\subsection{$P_{11}(1440)$ resonance}     
The first comprehensive $P_{11}(1440)$ electrocoupling sets available from 
the CLAS data on
$N\pi$ and $\pi^+\pi^-p$ electroproduction provide
access to the active degrees of freedom in the $P_{11}(1440)$ structure 
at various distances. 
Previous studies  \cite{AznRoper,Az09}  already showed a strong sensitivity of the electrocouplings
to assumptions
about the active components contributing to the $P_{11}(1440)$ structure.

The $P_{11}(1440)$ resonance is characterized by peculiar features:
 \begin{itemize}
\item  a smaller mass (1440 MeV) than that expected by quark models with 
vector flavor-conserving 
quark-quark ($qq$) interactions, this is less 
than the mass of the first orbital excitation of the nucleon, the
$S_{11}(1535)$, leading to so-called wrong $P_{11}(1440)$/$S_{11}(1535)$ mass
ordering \cite{Bu11b}, and 
\item a large hadronic decay width of $\approx$ 300 MeV, although other resonances with masses 
below 1.6 GeV have decay
widths of less than 200 MeV,
\end{itemize}
that make it difficult 
to describe the $P_{11}(1440)$ as a pure three quark bound state. 
To the best of our knowledge, 
there are no models, incorporating only  quark degrees of freedom, 
capable of describing simultaneously the mass ordering  
of the $P_{11}(1440)$ and $S_{11}(1535)$ 
resonances, the unusually large $P_{11}(1440)$ hadronic decay width, and 
the photo and electrocouplings. Therefore, different
approaches for the description of the $P_{11}(1440)$ were developed that 
take into
account the contributions from degrees of freedom 
other than those of dressed quarks \cite{Lee10,Lee08,Li92,Ob11}. The CLAS results on the $P_{11}(1440)$
electrocouplings of \cite{Az09} and this analysis allows for the first 
time to confront the model expectations for the $P_{11}(1440)$ structure 
with the experimental data, in particular
at $Q^2$ $<$ 0.6 GeV$^2$, where consistent results 
from the independent analyses of $N\pi$
and $\pi^+\pi^-p$ electroproduction off protons have now 
become available.

In Ref. \cite{Li92} the $Q^2$-dependence of the $P_{11}(1440)$ electrocouplings was obtained by assuming a hybrid 
nature of this
resonance. In this approach, the $P_{11}(1440)$  is treated as a resonance 
dominated by a single SU(6) configuration of three  dressed quarks 
oscillating against an explicitly excited glue, denoted $q^{3}G$.
 The $P_{11}(1440)$ as a hybrid state has a spin-flavor part of the wave function orthogonal 
to that of the nucleon, whereas the spin-flavor part of wave
function for the $P_{11}(1440)$, as a radially excited three quark state,  
is identical to that of the nucleon. This difference makes it possible
to distinguish between the $P_{11}(1440)$ as a regular 
$q^{3}$ or hybrid $q^{3}G$ state by studying  the 
$Q^2$-evolution of the $P_{11}(1440)$ electrocouplings. A characteristic prediction 
of \cite{Li92} for the hybrid origin 
of the $P_{11}(1440)$ is the absence of the longitudinal coupling, i.e.
$S_{1/2}=0$. This is a consequence of
negligible SU(6) configuration mixing in the non-relativistic 
approximation adopted in \cite{Li92} for the description of the $P_{11}(1440)$ 
structure. The recent Lattice QCD (LQCD) studies of the $N^*$ spectrum \cite{Ed11,Ed12} also confirmed the presence of the leading SU(6)-spin-flavor-configuration in the structure of $P_{11}(1440)$. The $S_{1/2}=0$ values for $P_{11}(1440)$ are clearly in disagreement 
with the results of our
analysis of the $\pi^+\pi^-p$ electroproduction data (see Fig.~\ref{p11_23}),
showing that at small photon virtualities $S_{1/2}$ is
larger than $A_{1/2}$, or at least comparable at $Q^2$ close to zero. 
This allows us to rule out sizable hybrid contributions to the 
$P_{11}(1440)$, confirming the conclusions 
from our previous analysis of
the $N\pi$ final state \cite{Az09}. The lack of a substantial contribution of a hybrid component to the lowest mass resonances in the $P_{11}$ partial wave was also confirmed in the LQCD studies \cite{Ed12}.

\begin{figure*}[htp]
\begin{center}
\includegraphics[width=8.8cm]{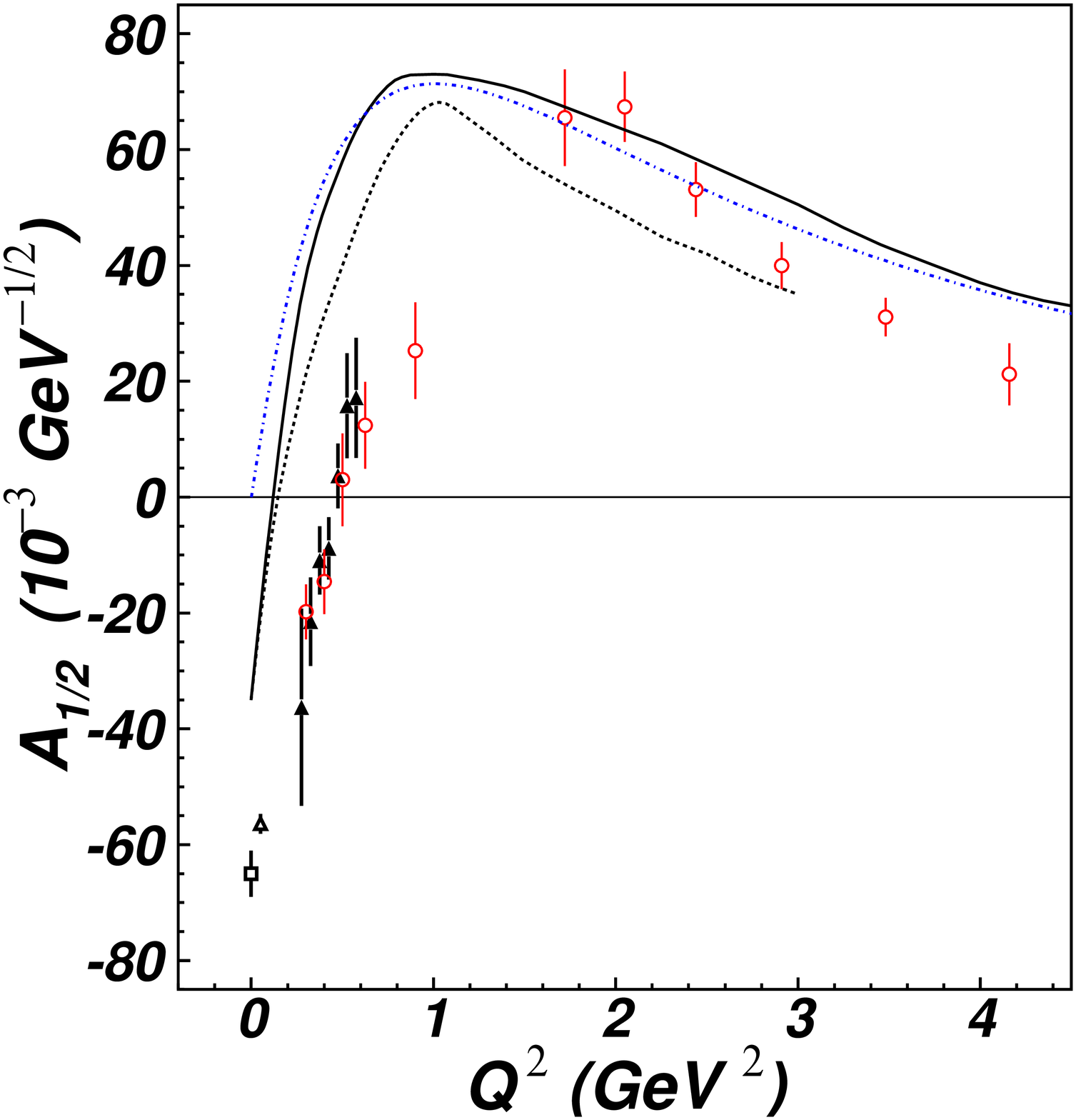}
\includegraphics[width=8.8cm]{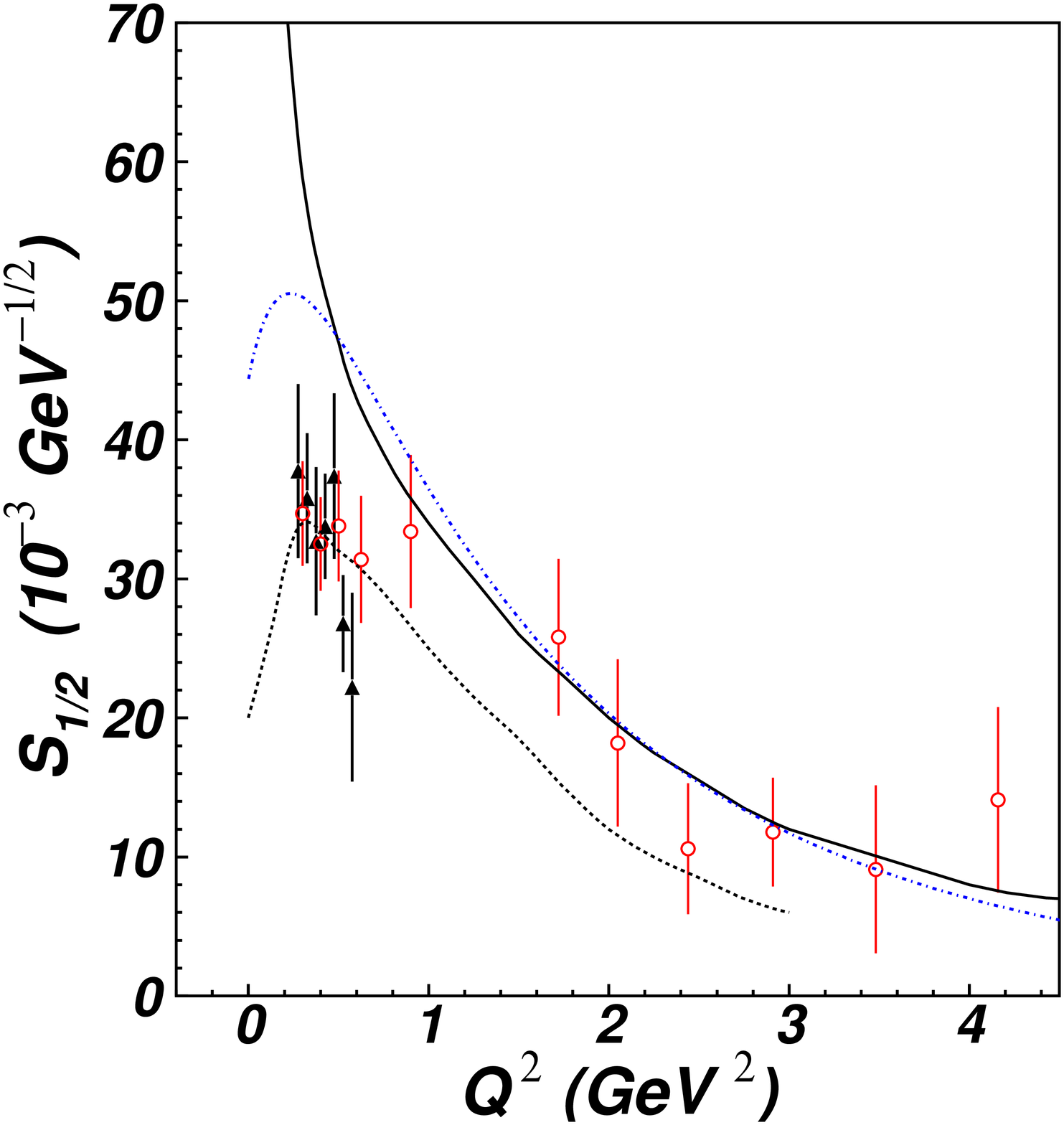}

\caption{\small(color online) Photo- and electrocouplings of the $P_{11}(1440)$ resonance in comparison with
predictions of quark models. The photocouplings are taken from RPP \cite{PDG10} (open square) and the CLAS data analysis \cite{Dug09} (open triangle). Other data points 
are the same as in Fig.~\ref{p11_23}. The results of relativistic light-front quark models 
\cite{AznRoper,Capstick} are shown by solid and dashed lines, respectively. Results of the covariant valence quark spectator
diquark model \cite{Ra10} are shown by the dashed dotted line.}  
\label{lf_p11}
\end{center}
\end{figure*}

The previous analysis \cite{Az09} of the $N\pi$ CLAS data 
showed that the most 
satisfactory description of the $P_{11}(1440)$ electrocouplings 
was achieved at $Q^2$ $>$ 1.5 GeV$^2$ within the framework 
of relativistic light-front quark models \cite{Capstick,AznRoper}, as shown 
in Fig.~\ref{lf_p11}. 
In these models, the $P_{11}(1440)$ electrocouplings are evaluated from 
the fully
covariant, one-body electromagnetic transition current for point-like 
constituent quarks in the impulse approximation, 
employing light-front relativistic
Hamiltonian dynamics \cite{Dr70,Be77,Ke91}. 
Resonance electrocouplings are computed 
in a frame where the ``+" component of the 
virtual photon light-cone momenta is equal to zero.
For this particular choice, the contribution from $q\bar{q}$ pairs into 
photon propagation are eliminated \cite{AznRoper}. 
In these models \cite{Capstick,AznRoper} 
the $P_{11}(1440)$ is
treated as the first radial excitation of three constituent quarks. 
The wave function of the ground state 
and $P_{11}(1440)$ are described under the simplifying assumption, 
that they are unmixed single 
oscillator basis states.

A reasonable description of the CLAS data 
at $Q^2$ $>$ 1.5 GeV$^2$ is achieved in both models \cite{Capstick,AznRoper},
as seen in Fig.~\ref{lf_p11}.  Accounting for the relativistic transition operator 
is important in
order to describe the CLAS  data over the entire range of photon virtualities, 
and especially at small $Q^2$ where the sign  
of $A_{1/2}$ changes \cite{Capstick,AznRoper}, although a quantitative 
 description of the data at small  $Q^2$ has not been achieved (see Fig.~\ref{lf_p11}).

A similarly good description of the $P_{11}(1440)$ electrocouplings at 
$Q^2$ $>$ 1.5 GeV$^2$ was 
found using a valence quark 
model \cite{Ra10} based on the covariant spectator formalism \cite{Gr08,Gr92}. The results of this model 
are also shown in Fig.~\ref{lf_p11}. Incorporating three constituent quarks only, this model treats the $N^*$
electroexcitation as a virtual photon interaction with a valence quark in the ground state, while the two other
quarks are combined to a spectator diquark in both spin states 0 and 1. 
The wave function of the ground
state is parametrized and fit to the data on elastic nucleon form factors. 
The $P_{11}(1440)$ structure is described as the first radial excitation of the nucleon ground state quarks. 
With no additional parameters,
the wave function of the $P_{11}(1440)$ was calculated employing the orthogonality condition 
between the ground state and 
the $P_{11}(1440)$ wave functions. The electrocouplings of $P_{11}(1440)$ are calculated for 
the valence quark transition between the nucleon ground state and its first radial excitation.
As shown in Fig.~\ref{lf_p11}, this model \cite{Ra10} describes the trend of the  CLAS data for 
$Q^2$ $>$ 1.5 GeV$^2$.

Different approaches to describe the $P_{11}(1440)$ resonance using structureless constituent quarks of non-running masses -- two relativistic light front quark models \cite{Capstick,AznRoper} and a valence quark model \cite{Ra10} -- are capable of reproducing the major features of
the CLAS data on 
$P_{11}(1440)$ electrocouplings only for 
$Q^2$ $>$ 1.5 GeV$^2$. These approaches fail 
to describe the electrocouplings for $Q^2$ $<$ 1.0 GeV$^2$. On the other
hand, consistent results 
on $P_{11}(1440)$ electrocouplings from the independent analyses
of $N\pi$ and $\pi^+\pi^-p$ electroproduction off protons at $Q^2$ $<$ 0.6
GeV$^2$ emphasize the credibility of the experimental data  and the inability of  
describing the $P_{11}(1440)$ structure by only accounting for quark  
degrees of freedom without meson-baryon dressing.

\begin{figure}[htp]
\begin{center}
\includegraphics[width=8cm]{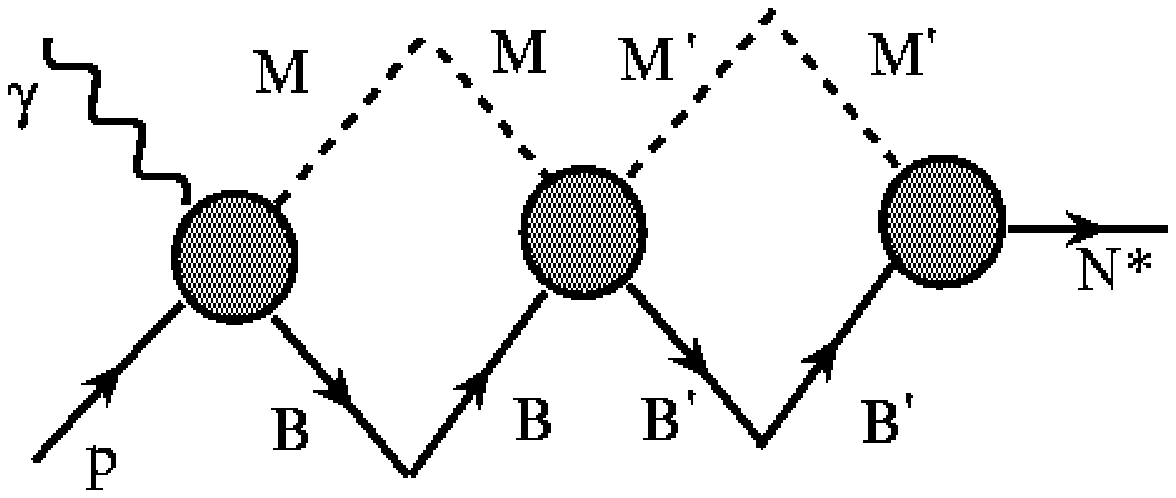}
\caption{\small Processes that contribute to meson-baryon dressing in the $N^*$ photo/electroexcitations within the EBAC-DCC
approach.}  
\label{mb_dress}
\end{center}
\end{figure}

The global analysis of $N\pi$, photo-, electro-, and hadroproduction within the framework of  
the EBAC-DCC coupled-channel approach \cite{Lee96,Lee01,Lee07,Lee08,Lee09,Lee09j,Lee10}  revealed
this additional component in the $N^*$ structure that is most relevant 
at $Q^2$ $<$   1.0
GeV$^2$.   
The general unitarity condition for full meson electroproduction amplitudes 
requires contributions from
non-resonant meson electroproduction and hadronic scattering amplitudes 
to the $\gamma_{v}pN^*$ vertex, as depicted in Fig.~\ref{mb_dress}. This contribution incorporates
all possible intermediate meson-baryon states and subsequent meson-baryon 
scattering processes
that eventually result in the $N^*$ formation. 
As was mentioned in Section~\ref{nstarelectrocoupl}, the $\gamma_{v}pN^*$ electrocouplings 
extracted in our
analysis of $\pi^+\pi^-p$ electroproduction data, as well as in the previous
analysis of the $N\pi$ data \cite{Az09} account for $all$ contributions to
the resonance structure, including for those from meson-baryon dressing. 
Instead, most of quark models, including aforementioned \cite{Capstick,AznRoper,Ra10}, account only for the quark-core component in $N^*$ photo-/electroexcitation, while 
meson-baryon dressing, being a part of the reaction mechanism, is completely outside of their 
scope. Therefore, the current and the previous CLAS results \cite{Az09} on $\gamma_{v}pN^*$ electrocouplings 
should be different with respect to those from the quark model expectations 
in $Q^2$-areas where the contributions 
from meson-baryon dressing  are substantial.

In Ref.~\cite{Lee08} the meson-baryon dressing amplitudes 
are calculated from diagrams describing 
non-resonant mechanisms in $N\pi$ photo-, electro-, and 
hadroproduction channels with hadronic parameters 
fit to the world meson hadroproduction data.   
The estimates of the absolute values for meson-baryon dressing
contributions to the $P_{11}(1440)$ electrocouplings showed that meson-baryon
dressing amplitudes are maximal at $Q^2$ $<$ 1.0 GeV$^2$. 

Therefore the meson-baryon dressing contributions could be  responsible 
for the differences observed in this area 
between the CLAS data on the $N^*$ electrocouplings 
and the quark model expectations of Refs. \cite{Capstick,AznRoper,Ra10}. 
The meson-baryon dressing
decreases with $Q^2$ and in the region $Q^2$ $>$ 1.0 GeV$^2$ we have 
a gradual transition to the dominance of
quark degrees of freedom, as indicated by the better 
description of the $P_{11}(1440)$
electrocouplings within the framework of quark models. 

Evaluations for bare-quark-core contributions to the dressed $P_{11}(1440)$ electrocouplings have recently be obtained within the framework
of QCD-based Dyson-Schwinger Equations (DSEQCD) \cite{CR2012}. The DSEQCD results
are close to the bare-quark-core contributions to the $P_{11}(1440)$ 
electrocouplings extracted from the experimental
data within the framework of the EBAC-DCC coupled-channel approach \cite{Lee10} 
and far from the dressed $P_{11}(1440)$ electrocouplings obtained in our analysis. 
This DSEQCD result offers further evidence for substantial contributions 
from meson-baryon dressing to the $P_{11}(1440)$ electrocouplings.

We conclude that the structure of $P_{11}(1440)$ is determined by the 
combined
contributions from an internal quark core of constituent quarks in the
first radial excitation 
and an external meson-baryon dressing, 
the latter being most relevant at small $Q^2$.


Lattice QCD is making progress in calculating $\gamma_{v}NN^*$ 
transition form factors 
from first QCD
principles. Exploratory calculations of the $\gamma_{v}pP_{11}(1440)$ transition form factors $F^{P11}_{1,2}(Q^2)$ 
were carried out within LQCD in the quenched approximation using a pion mass of 500 MeV and a simplified set of projection
operators \cite{Li09,Li09a}. Despite the aforementioned simplifications, 
the first LQCD evaluations are in approximate agreement with the 
CLAS $F^{P11}_{1,2}(Q^2)$ form factor results
at  $Q^2$ $>$ 1.5 GeV$^2$. Improved LQCD calculations with dynamical 
quarks and a smaller pion mass of 380 MeV \cite{Li11,Li11a} provide for the first time a reasonable description of the CLAS data 
for $Q^2$ $<$ 1.0 GeV$^2$. These studies demonstrated promising opportunities 
to relate the results 
on the $Q^2$-evolution of the $\gamma_{v}pN^*$ electrocouplings 
directly to QCD. 

Together with the recent LQCD  \cite{Ed11,Ed12} and DSEQCD \cite{CRo103}
results on the $N^*$ spectrum and the first DSEQCD results on bare
$\gamma_{v}pP_{11}(1440)$ electrocouplings \cite{CR2012}, this is an important step
toward to our understanding of baryon formation 
and the nature of confinement from the first principles of QCD.

The $\pi \Delta$ and $\rho p$ $P_{11}(1440)$ hadronic decay widths obtained 
in our
analysis are listed in Table~\ref{hpp11} and compared with RPP \cite{PDG10} values. The
quark models, which describe $N^*$  hadronic decays through  flux tube 
breaking \cite{St89,St95}, provide a 
good description of our $\pi \Delta$ and 
$\rho p$ decay widths of 100 and  2.5 MeV, respectively. However, the flux-tube breaking mechanism is unable to 
describe the $P_{11}(1400)$ $N\pi$  hadronic decays. It  predicts a width of 
approximately 400
MeV for this channel, which is in strong disagreement with measured values \cite{PDG10}.  
Other approaches, listed in Ref.~\cite{Ca00}, fail to describe the $N\pi\pi$ hadronic decay widths.

To date the quantitative description of 
$P_{11}(1440)$ hadronic decays remains
a challenging problem. Consistent  accounting for the meson-baryon  dressing contributions to 
both electromagnetic and  
hadronic $P_{11}(1440)$ vertices is important to gain insight into the structure 
of this excited state \cite{Lee101,Lee101a,Ob11,CR2012}. 

\subsection{$D_{13}(1520)$ resonance}

\begin{figure}[htp]
\begin{center}
\includegraphics[width=7cm]{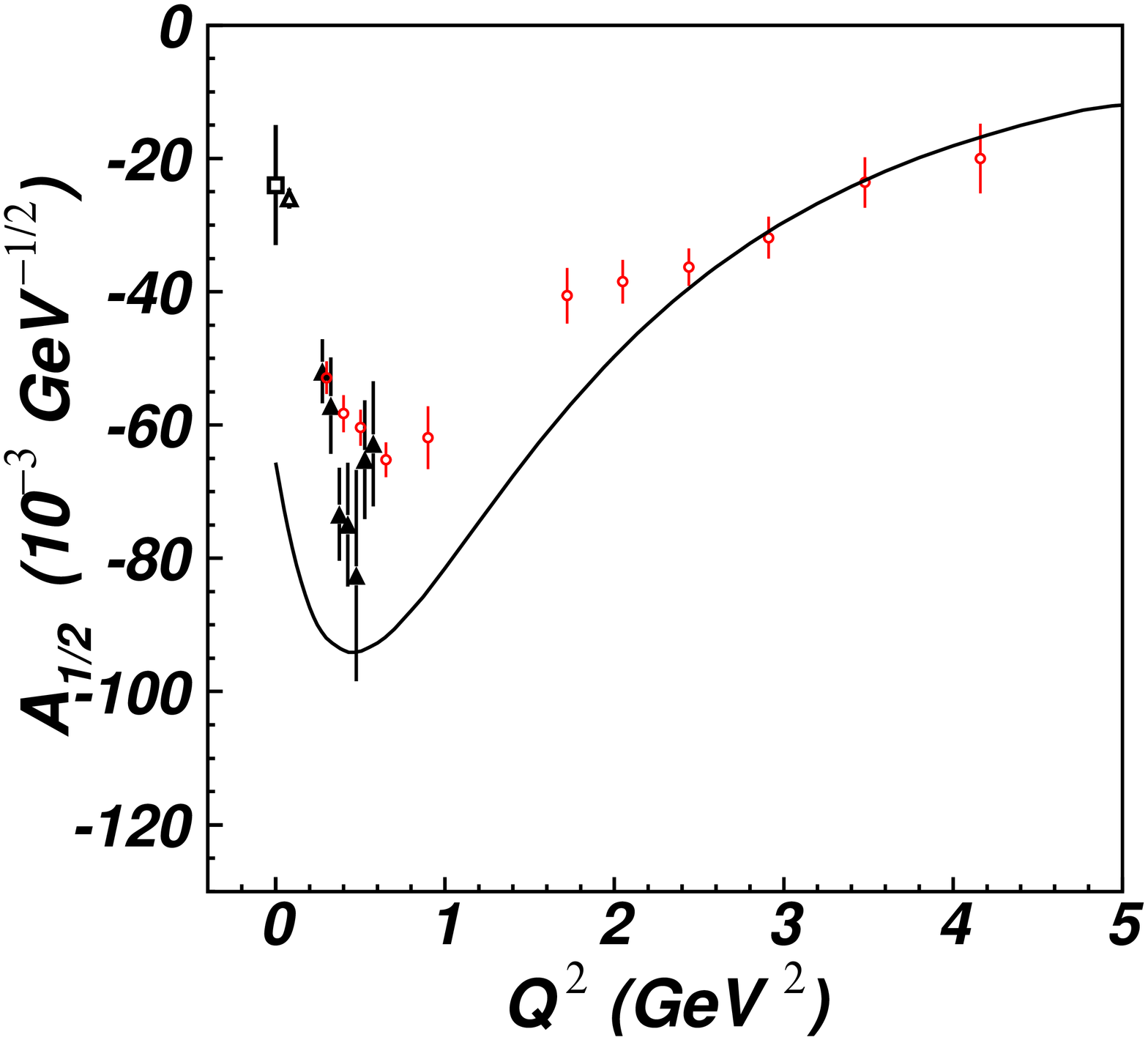}
\includegraphics[width=7cm]{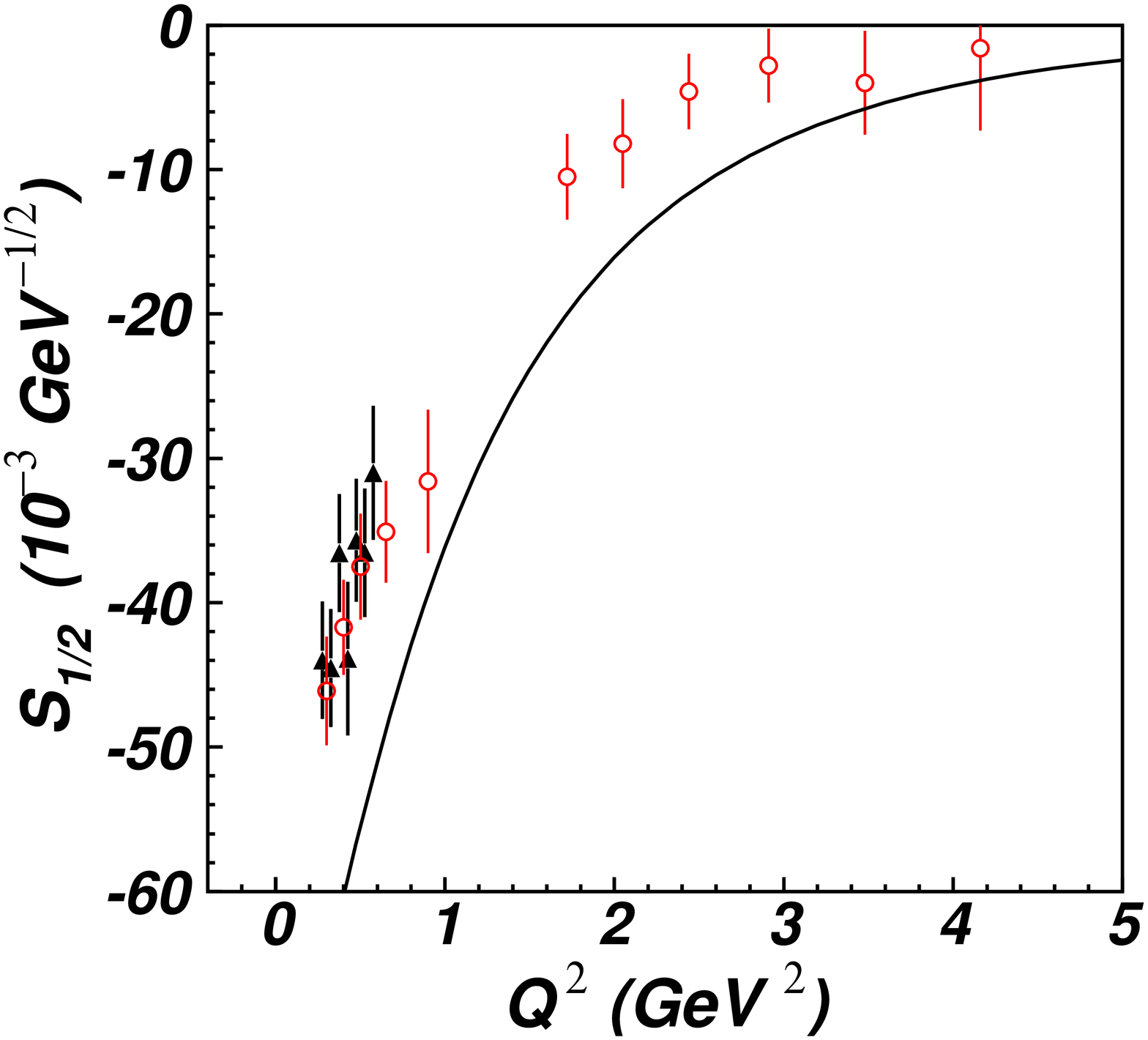}
\includegraphics[width=7cm]{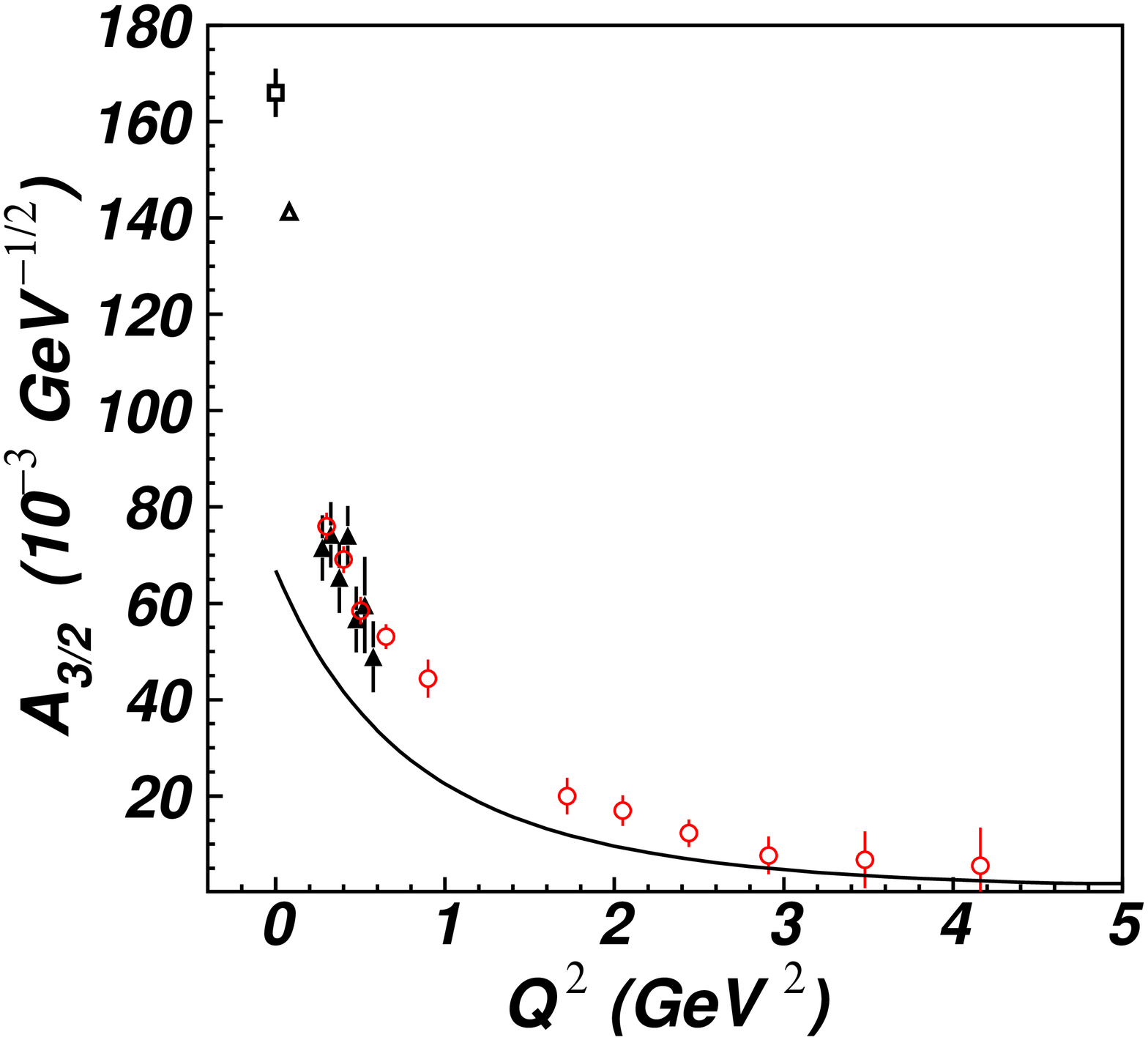}
\caption{\small(color online) Description of the CLAS data on $D_{13}(1520)$ 
electrocouplings
within the hQCM (solid lines) \cite{Gia98}. The symbols associated with
 the experimental 
data are the same as in Fig.~\ref{lf_p11}.}  
\label{d13_hcqm}
\end{center}
\end{figure}

Analyses of the previous CLAS data \cite{Az09} on $D_{13}(1520)$
electrocouplings determined from $N\pi$ exclusive electroproduction channels,
showed that their satisfactory description for 
$Q^2$ $>$ 1.5 GeV$^2$ can be achieved within the framework of
the hypercentral constituent quark model (hCQM) \cite{Gia98} 
(solid lines in Fig.~\ref{d13_hcqm}). In this model the central confinement potential 
is parametrized by a sum of Coulomb and linear terms expressed in collective
hypercoordinates for the three constituent quark system. The use of 
hypercoordinates effectively accounts for three-body effects in quark interactions.
The remaining interaction between pairs of quarks is parametrized by a superposition of spin- 
and isospin-dependent hyperfine terms. Wave functions for resonances are
obtained by diagonalizing the Hamiltonian in a non-relativistic approximation. 
Three parameters of the hQCM were fit to data on 
the baryon spectrum. With these parameters
electrocouplings of all well-established excited nucleon states 
were evaluated for $Q^2$ $<$ 5.0
GeV$^2$, employing non-relativistic electromagnetic transition operators. The results obtained
for the $D_{13}(1520)$ state are shown in Fig.~\ref{d13_hcqm}.

The hCQM model can describe the data trends for the $D_{13}(1520)$ 
electrocouplings for $Q^2$ $>$ 1.5 GeV$^2$. Pronounced differences between 
hCQM expectations and
the $D_{13}(1520)$ electrocouplings at smaller $Q^2$ offer an indication for
contributions from active degrees of freedom other than a core of three dressed quarks to this state's
electrocouplings. Meson-baryon dressing amplitudes are likely contributors in
this area of photon virtualities. According to the EBAC analysis \cite{Lee08}, for the $D_{13}(1520)$ state 
they are maximal 
for small photon virtualities and 
decrease with $Q^2$. The EBAC analysis also suggests negligible meson-baryon dressing or dominant quark core 
contributions to the $A_{1/2}$ 
electrocoupling of the $D_{13}(1520)$ state for $Q^2$ $>$ 1.5 GeV$^2$, where we already have the CLAS results on 
this electrocoupling  \cite{Az09}. 
These results thus offer access to quark degrees of freedom 
in the structure of $D_{13}(1520)$ and open up new prospects to study the dynamical 
dressed quark  mass, structure, 
and their strong interactions, that are
responsible for the  $N^*$ formation. These studies are of particular 
importance to understand 
the nature of confinement in the baryon sector based 
on QCD \cite{CRo101,CRo103,CRo104,Li09,Li11,Br09,Brod11,CL12}.

\begin{figure}[htp]
\begin{center}
\includegraphics[width=8cm]{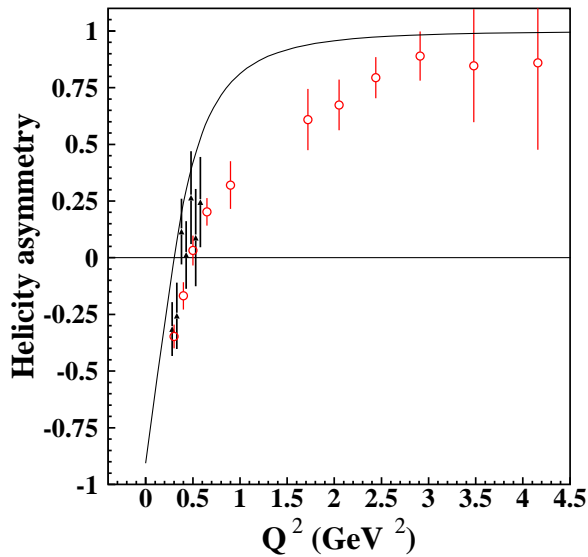}
\caption{\small(color online) Helicity asymmetry $A_{hel}$ of 
Eq.~(\ref{held13form}) 
for the transverse electrocouplings of the 
$D_{13}(1520)$ from this analysis of $\pi^+\pi^-p$ (filled
triangles) and $N\pi$ (open circles) exclusive electroproduction channels measured with CLAS. 
The curve
represents the non-relativistic quark model calculations \cite{Ko80}.}  
\label{held13}
\end{center}
\end{figure}

Our analysis confirms  a rapid helicity switch from the dominance of the $A_{3/2}$ 
electrocoupling at the photon point to a comparable contribution from both transverse electrocouplings 
at $Q^2$ $\approx$ 0.5 GeV$^2$ as already observed in \cite{Az09}.
This is shown in Fig.~\ref{held13} in terms of the
helicity asymmetry, defined as
\begin{equation}
\label{held13form} 
A_{hel}=\frac{A^{2}_{1/2}-A^{2}_{3/2}}{A^{2}_{1/2}+A^{2}_{3/2}} .
\end{equation}
This particular feature is expected for the contributions from the core of three constituent quarks 
in the first orbital nucleon excitation $L=1$ \cite{Ko80}. 
It suggests a significant 
contribution  from the core of three constituent quarks  to 
the transverse $D_{13}(1520)$ electrocouplings even at small $Q^2$.

We conclude that the $Q^2$-evolution of the $D_{13}(1520)$  electrocouplings is consistent with 
contributions of both an external meson-baryon cloud and an internal core of three constituent quarks.

The $N\pi\pi$ hadronic decay widths of $D_{13}(1520)$, obtained from 
the CLAS $\pi^+\pi^-p$
data, are given in the Table~\ref{hpd13}. They are close to the 
RPP \cite{PDG10} values. However, the branching fraction to the $\pi \Delta$
final state is larger  and to the $\rho p$ final state is smaller 
than the RPP average. 
Our results on the $N\pi\pi$ hadronic couplings of the $D_{13}(1520)$ are well described by the 
flux tube breaking model of \cite{St89,St95}, as well
as within the framework of the $^{3}P_{0}$ pair creation model of \cite{Ca94}. The success of these models 
indicates a substantial role of the quark core in hadronic decays of the $D_{13}(1520)$ state. 

\section{Summary and outlook}
 The analysis of a large body of differential and fully integrated cross sections for 
the process $\gamma_{v}p \rightarrow \pi^+ \pi^- p$ carried out within 
the framework of the JM model in our previous paper \cite{Mo09}
allowed us to establish all essential mechanisms that contribute 
to this exclusive channel for 
1.3 GeV $<$ $W$ $<$ 1.6 GeV and 0.2 GeV$^2$ $<$ $Q^2$ $<$ 0.6 GeV$^2$. 
In this paper we use
the reaction model of Ref. \cite{Mo09} in order to determine the resonant and
non-resonant contributions to the $\pi^+\pi^-p$ differential cross sections
measured with the CLAS detector \cite{Fe09} and eventually to obtain 
the $A_{1/2}$, $S_{1/2}$, and $A_{3/2}$ electrocouplings, as well as the 
$\pi \Delta$ and $\rho p$ decay widths for the $P_{11}(1440)$ and $D_{13}(1520)$ excited proton 
states from this
data fit.   

The good description of all measured 
observables makes it possible to evaluate the resonant contributions 
to the cross sections, which are needed to extract the $\gamma_{v}pN^*$
electrocouplings. Resonance electrocouplings and  hadronic decay parameters were determined
using a unitarized BW ansatz for the resonant amplitudes, that takes into account 
interactions between the same and different excited states in the dressed resonance propagators.

For the first time electrocouplings of the $P_{11}(1440)$ and $D_{13}(1520)$ resonances were obtained 
from the analysis of the $\pi^+\pi^-p$ exclusive electroproduction at photon
virtualities 0.25 GeV$^2$ $<$ $Q^2$ $<$ 0.60 GeV$^2$. They are in 
reasonable agreement with
the electrocoupling values obtained in the previous CLAS analysis of the exclusive $N\pi$
electroproduction channels \cite{Az09}. 
Single and charged double pion electroproduction 
channels represent two major contributors to meson electroproduction in the
kinematical region covered by our measurements. The non-resonant mechanisms 
in these exclusive channels are
completely different. 
A successful description of a large body of observables in single and charged 
double pion
electroproduction channels with consistent values of  the $P_{11}(1440)$ and 
$D_{13}(1520)$ electrocouplings
confirms that the $\gamma_{v}pN^*$ electrocouplings can be reliably determined in independent analyses
of these electroproduction channels.
A good agreement between electrocouplings of the aforementioned excited states obtained 
from $N\pi$ and $\pi^+\pi^-p$
electroproduction also demonstrate that the reaction models developed 
to describe 
these exclusive channels \cite{Az09,Mo09} can be used to determine 
electrocouplings  of most of the excited proton states that decay preferentially 
into either $N\pi$ or $N\pi\pi$ final states. 

The $\pi \Delta$ and $\rho p$ partial 
hadronic decay widths of
the $P_{11}(1440)$ and 
$D_{13}(1520)$ states were also determined. They are close to 
the results of experiments with hadronic probes reported in the RPP \cite{PDG10}.

The comparison of the CLAS data on $P_{11}(1440)$ and $D_{13}(1520)$ electrocouplings 
with the expectations of conceptually different  
quark models \cite{Capstick,AznRoper} and \cite{Ra10}, complemented by evaluations 
of meson-baryon dressing \cite{Lee08} allowed us to shed light on the relevant components 
in the structure of these states and to explore their
evolution with photon virtualities. The electrocouplings reveal
two major contributions, from an internal core of three dressed quarks
and an external meson-baryon cloud. 

The structure of the $P_{11}(1440)$  is consistent 
with a combined contribution from constituent quarks in the first 
radial excitation and  from a meson-baryon
cloud with pronounced contributions  at $Q^2$ $<$ 0.6 GeV$^2$.
The electrocouplings of the $D_{13}(1520)$ contain a combined 
contribution of three constituent quarks 
in the first orbital nucleon excitation $L=1$ and an external 
meson-baryon cloud, which becomes negligible for
the $A_{1/2}$ electrocoupling of $D_{13}(1520)$ for $Q^2$ $>$ 1.5 GeV$^2$. 
The CLAS results on the $A_{1/2}$  electrocoupling of the $D_{13}(1520)$ resonance 
in this kinematical region provide direct 
access to quark degrees of freedom 
in the structure of this excited state. A physical interpretation of these results is of particular 
interest for those who are studying hadron structure starting from the QCD Lagrangian \cite{CL12,CR2012,Br09,Li09,Li11,Li11a,Ed11,Ed12,Brod11}.

The resonance electrocouplings and hadronic decay parameters presented 
in this paper were 
obtained using the unitarized BW parametrization of
resonant amplitudes. Their values are determined at the resonant point 
on the real energy axis ($W=M_{r}$) and incorporate $all$ combined relevant
contributions (quark-core, meson-baryon dressing and so on). 
The $N^*$ parameters extracted from the data in this way  can only be compared 
directly with those obtained from the same or 
other exclusive channel data fits that 
employ BW parametrizations of resonant amplitudes. 
It remains to be seen how these resonance
parameters can be related to the values determined  
from singularities of the reaction amplitudes in the complex energy plane, as employed in 
coupled-channel 
analyses \cite{Lee10}.
In the Section~\ref{regbw} we described the JM model relations between 
the resonance parameters and the model independent observables: the resonant 
part of the cross sections and 
$N^*$ electromagnetic/hadronic decay widths. We can use the model
independence of these observables and require agreement of the cross section and/or resonance decay widths
at the resonant point evaluated in different models. These observables, in turn, could be related to the resonance 
parameters for any particular model. 
This should make it possible to establish
relations between resonant parameters determined in
our approach and those from a global multi-channel analyses.

\section{Acknowledgments}
We would like to acknowledge the outstanding efforts of the staff of the Accelerator and the Physics Divisions at Jefferson Lab that made this evaluation of $P_{11}(1440)$ and  
$D_{13}(1520)$ electrocouplings possible. We are grateful to I. J. R. Aitchison, I. G. Aznauryan, R. G. Edwards, M. M. Giannini, T.-S. H. Lee, H-W. Lin, C. D. Roberts, and E. Santopinto for helpful discussions. This work was supported
in part by the U.S. Department of Energy and the National
Science Foundation, the Russian Federation
Government Grant 02.740.11.0242, 07.07.2009, the Skobeltsyn Institute of Nuclear Physics and
Physics Department at Moscow State University, University of South Carolina, Yerevan Physics Institute (Armenia), the Chilean Comisi\'on Nacional de Investigaci\'on Cient\'ifica y Tecnol\'ogica (CONICYT), the French Centre National de la Recherche Scientifique (CNRS), the French Commissariat a l'Energie Atomique, the Italian Istituto Nazionale di Fisica Nucleare, the National Research Foundation of Korea,
and the UK Science and Technology Facilities Research Council (STFC), the Scottish Universities Physics Alliance (SUPA), and the United Kingdom's Science and Technology Facilities Council. The Southeastern Universities Research Association (SURA) operates the
Thomas Jefferson National Accelerator Facility for the United States
Department of Energy under contract DE-AC05-84ER40150.

\include{appendix1}
\end{document}

%% file: appendix1.tex
\appendix

\section*{Appendix A: Resonance hadronic decay amplitudes employed in the JM model
\label{hadrwidths}}

The relationship within the JM model \cite{Mo09,Ri00} between the $N^*$ hadronic decay amplitudes  
$\langle \lambda_{f} \vert T_{dec} \vert 
\lambda_{R} \rangle$ and the energy-dependent $N^{*}$ partial hadronic decay widths   
$\Gamma_{\lambda_{f}}(W)$ is presented in this Appendix.  Here $\lambda_R$ is  
the helicity of the $'$N*$'$, which decays into $\pi\Delta$ or $\rho p$ final states with  
helicity $\lambda_{f}$.

 The $N^*$ hadronic decay amplitudes in Eq. (\ref{regbwamp}) can be expanded in partial waves of total 
 momentum $J$: 
\begin{eqnarray}
\label{angdecompa} 
\langle \lambda_{f} \vert T_{dec} \vert \lambda_{R} \rangle=\sum_{J} \langle \lambda_{f} \vert T^{J}_{dec} \vert 
\lambda_{R} \rangle d^{J}_{\mu\nu}(\cos\theta^{*})e^{i\mu\phi^{*}},  
\end{eqnarray}
where $\theta^{*}$ and $\phi^{*}$ are the CM emission angle of the $\pi$ 
for the $\pi \Delta$
intermediate state and of the final $'$p$'$ for the $\rho$ p intermediate states. The indexes $\mu$ and $\nu$ are defined in Eq. (\ref{decaywidth5}). Only a single term 
with $J$=$J_{r}$, where $J_{r}$ is resonance spin,
contributes to the expansion in Eq. (\ref{angdecompa}) because of 
total angular momentum conservation.
We can rotate the quantization axis adopted for the initial-state 
$\vert \lambda_{R} \rangle$ wave function and 
re-evaluate the matrix element $\langle \lambda_{f} \vert T_{dec} \vert \lambda_{R} \rangle$ 
in the frame with the quantization axis defined by the final 
$\pi$  (the final $'$p$'$) 
momentum for $\pi \Delta$ ($\rho p$) $N^*$ 
 decays, respectively. The matrix element $\langle \lambda_{f} \vert T_{dec} \vert \lambda_{R} \rangle$
after rotation becomes:
\begin{eqnarray}
\label{angdecomp1} 
\langle \lambda_{f} \vert T_{dec} \vert \lambda_{R} \rangle=\sum_{\nu'}\langle \lambda_{f} \vert T_{dec} \vert 
J_{r} \; \nu' \rangle d^{J_{r}}_{\mu\nu'}(cos\theta^{*})e^{i\mu\phi^{*}}. 
\end{eqnarray}
The superposition of the states $\vert 
J_{r} \; \nu' \rangle$ in Eq. (\ref{angdecomp1}), with spin $J_{r}$ and projection 
onto the final-state  quantization axis $\nu'$,  is the transformed wave function of the initial-state $\vert \lambda_{R} \rangle$ after the aforementioned rotation of the initial-state quantization axis.
Rotational invariance of the resonance hadronic decay amplitudes requires that the operator $T_{dec}$ should 
be an SU(2) $\otimes$ O(3) - spin $\otimes$ orbital momentum scalar.  Only the term 
with $\nu'=\nu$ in Eq. (\ref{angdecomp1}) 
(with $\nu$ defined by Eq. (\ref{decaywidth5})) 
remains non-zero in the sum of Eq. (\ref{angdecomp1}), as  a consequence of the Wigner-Eckart theorem 
applied to matrix elements $\langle \lambda_{f} \vert T_{dec} \vert 
J^{r} \; \nu' \rangle$ with the scalar $T_{dec}$ operator. 

From comparisons between Eqs. (\ref{angdecompa}) and (\ref{angdecomp1}) we can see that:
\begin{eqnarray}
\label{equal} 
\langle \lambda_{f} \vert T^{J_{r}}_{dec} \vert \lambda_{r} \rangle=\langle \lambda_{f} \vert T_{dec} \vert J^{r} 
\nu \rangle , 
\end{eqnarray} 
The $\langle \lambda_{f} \vert T^{J_{r}}_{dec} \vert 
\lambda_{R} \rangle$ matrix elements in Eqs. (\ref{decaywidth5},\ref{equal}) are determined by the final-state helicity 
$\lambda_{f}$ only, and  
are independent 
of $N^*$ helicities $\lambda_{R}$. 

  
Assuming real values for the matrix element $\langle \lambda_{f} \vert T^{J_{r}}_{dec} \vert 
\lambda_{r} \rangle$ 
in Eqs.~(\ref{decaywidth5},\ref{equal}), we can relate it to the $\Gamma_{\lambda_{f}}(W)$ 
partial hadronic decay width of the excited state $N^{*}$ to the final state of helicity
$\lambda_{f}$. We employ general relations of quantum theory 
between 
the resonance decay amplitude $\langle \lambda_{f} \vert T_{dec} \vert 
\lambda_{R} \rangle$ of Eq. (\ref{angdecompa}),
the two-body state phase space of resonance decay products $d\Phi_{2b}$, 
and  the $\Gamma_{\lambda_{f}}(W)$ decay width:
\begin{eqnarray}
\Gamma_{\lambda_{f}}(W)=\frac{1}{2M_{r}}\frac{1}{2J_{r}+1}\sum_{\lambda{R}}\int \vert\langle \lambda_{f} \vert T_{dec} \vert 
\lambda_{R} \rangle \vert^{2}d\Phi_{2b}.  
\label{ampwidth}
\end{eqnarray}
The factor $\frac{1}{2M_{r}}$ in Eq. (\ref{ampwidth}) reflects the spin-tensor normalization in the convention 
of the JM model \cite{Mo09}. This normalization, and the expression for the $'$S$'$-matrix adopted in the JM model 
\cite{Mo09}, defines the final-state two-body  phase space $d\Phi_{2b}$  as:
\begin{eqnarray}
d\Phi_{2b}=\frac{1}{4\pi^{2}}\frac{p_{f}}{4M_{r}}\sin(\theta^{*})d\theta^{*}d\phi^{*},  \\
E_{f}=\frac{W^2+M_{f}^2-M_{f'}^2}{2W}, \nonumber \\
p_{f}=\sqrt{E_{f}^2-M_{f}^2} \nonumber ,
\label{diff2b}
\end{eqnarray}
where $E_{f}$, $p_{f}$ are the energy and momentum modulus of one of the  final
hadrons $f$ ($f$ is either pion or the final proton for the $N^*$ $\rightarrow$ $\pi \Delta$ or $\rho$ $'$p$'$
decays, respectively), $M_{f}$ is its mass, 
while the index $f'$ stands for the other hadron. 
All frame-dependent kinematic variables of the final hadrons are defined 
in the  final hadron
CM frame.  Inserting Eqs. (\ref{angdecompa},\ref{equal}) into Eq. (\ref{ampwidth}) and accounting for and the $'$d-$'$function normalization,
\begin{equation}
\int d^{J*}_{\mu \, \nu}(\cos\theta^*)\cdot d^{J}_{\mu \, \nu}(\cos\theta^*) \sin\theta^* d\theta^*=\frac{2}{2J+1}, 
\end{equation}
we obtain Eq. (\ref{decaywidth5}) for
the $\langle \lambda_{f} \vert T_{dec} \vert \lambda_{R} \rangle$ amplitudes, apart from the factor  
$\sqrt{\frac{\langle p^{r}_{i} \rangle}{\langle p_{i} \rangle}}$. Note that at the resonant point, $W$=$M_{r}$,
this factor is equal to unity. However, in calculations of resonant cross sections for  
$W$$\ne$$M_{r}$, the two-body phase space becomes different than that at the resonant point. 
The factor $\sqrt{\frac{\langle p^{r}_{i} \rangle}{\langle p_{i} \rangle}}$ in Eq. (\ref{decaywidth5}) accounts for 
this difference. It is needed to evaluate resonant cross sections and amplitudes for $W$ values off 
the resonant point.

The $\langle p^{r}_{i} \rangle$ and $\langle p_{i} \rangle$  absolute 
three-momentum  values of the final $\pi$ for the   
$N^* \rightarrow \pi \Delta$ decay ($i$=1) or of the final $p'$ 
for the $N^* \rightarrow \rho p$ decay ($i$=2) in Eq. (\ref{decaywidth5}) are  
averaged over the running mass of the unstable hadron in the
intermediate state:

\begin{gather}
\label{averf}
\langle p^{r}_{i} \rangle=  
\int dM^{2}_{i}\frac{1}{\pi}\frac{M_{i\,0}\Gamma_{i\,0}}
{(M^{2}_{i}-M^{2}_{i\,0})^{2}+M^{2}_{0\, i}\Gamma^{2}_{0\,i}}p^{r}_{i}(M^{2}_{i}),  \\
\nonumber \\
\langle p_{i} \rangle=
\int dM^{2}_{i}\frac{1}{\pi}\frac{M_{i\,0}\Gamma_{i\,0}}
{(M^{2}_{i}-M^{2}_{i\,0})^{2}+M^{2}_{i\,0}\Gamma^{2}_{i\,0}}p_{i}(M^{2}_{i}) ,  \nonumber \\
\nonumber
\end{gather}
where $M_{i}$ is the current invariant mass of the final  
$\pi p$ particles in the case of $N^* \rightarrow \pi \Delta$
decay ($i=1$) or the current invariant mass of final $\pi\pi$ particles for the  
$N^* \rightarrow \rho p$ decay ($i=2$); $M_{i\,0}$ are the central masses of either $\Delta$ ($i=1$) or
$\rho$ ($i=2$); and $\Gamma_{i\,0}$, are their total decay widths.
The running momenta of the stable particles from $N^*$ decays  $p_{i}(M^{2}_{i})$ in 
Eq. (\ref{averf}) are evaluated as:
\begin{eqnarray}
\label{runmom}
p_{1}=\frac{[(W^2+m_{\pi}^{2}-M_{1}^2)^2-4W^{2}m_{\pi}^2]^{1/2}}{2W} \ \text{and} \\
\, p_{2}=\frac{[(W^2+m_{p'}^{2}-M_{2}^2)^2-4W^{2}m_{p'}^2]^{1/2}}{2W} . \nonumber
\end{eqnarray}
The values of these momenta at the resonance point $p^{r}_{i}(M^{2}_{i})$ were obtained from Eq. (\ref{runmom}) at  $W=M_{r}$.


The hadronic decay widths $\Gamma_{\lambda_{f}}(W)$ 
were taken from experiments
with hadronic probes, as was described in Section~\ref{fit}. Those decay widths were obtained in another
representation of orbital angular momentum $L$ 
and total final hadron spin $S$ $\Gamma_{LS_{f}}(W)$. 
The partial $\Gamma_{\lambda_{f}}(W)$ decay widths can be transformed into this 
representation \cite{Jac59} by:
\begin{eqnarray} 
\label{hadrwidth1}
\sqrt{\Gamma_{\lambda_{f}}}=\sqrt{\frac{2J_{r}+1}{2L+1}}\langle L 0 S \lambda_{1}-\lambda_{2} 
\vert J_{r} \lambda_{1}-\lambda_{2} \rangle \cdot \nonumber \\ 
\langle s_{1} \lambda_{1} s_{2} -\lambda_{2} \vert
S \lambda_{1}-\lambda_{2} \rangle\sqrt{\Gamma_{LS}}, 
\end{eqnarray}
\\
where $s_{1}$, $\lambda_{1}$, $s_{2}$, $\lambda_{2}$ are spins and helicities for first stable and
second unstable particles in the intermediate states.

\clearpage
\newpage